\documentclass[aps,pre,reprint,onecolumn,superscriptaddress]{revtex4}

\pdfoutput=1

\usepackage{helvet,times}
\usepackage{bm,textcomp}

\usepackage{amsmath}
\usepackage{ucs}
\usepackage[utf8x]{inputenc}
\usepackage[english]{babel}
\usepackage{color,graphicx}
\usepackage{amsmath,amsfonts,amssymb,color,fancybox}
\usepackage{units}
\usepackage{array,url,kantlipsum}
\usepackage{xcolor}
\usepackage{float}
\usepackage{soul}
\usepackage{mhchem}

\begin{document}

\setcounter{page}{1} 

\title{Drift and behavior of {\it E. coli} cells}

\author{Gabriele Micali}
\affiliation{Department of Life Sciences, Imperial College, London, United Kingdom}
\affiliation{Centre for Integrative Systems Biology and Bioinformatics, Imperial College, London, United Kingdom}
\affiliation{Department of Environmental Microbiology, Eawag, D\"{u}bendorf, Switzerland}
\affiliation{Department of Environmental Systems Science, ETH Z\"{u}rich, Z\"{u}rich, Switzerland}

\author{Remy Colin}

\author{Victor Sourjik}
\email{victor.sourjik@synmikro.mpi-marburg.mpg.de}
\affiliation{Max Planck Institute for Terrestrial Microbiology, Marburg, Germany}
\affiliation{LOEWE Center for Synthetic Microbiology, Marburg, Germany}

\author{Robert G. Endres}
\email{r.endres@imperial.ac.uk}
\affiliation{Department of Life Sciences, Imperial College, London, United Kingdom}
\affiliation{Centre for Integrative Systems Biology and Bioinformatics, Imperial College, London, United Kingdom}

\begin{abstract}%
Chemotaxis of the bacterium {\it Escherichia coli} is well understood in shallow chemical gradients, but its  swimming behavior remains difficult to interpret in steep gradients.  By focusing on single-cell trajectories from simulations, we investigated the dependence of the chemotactic drift velocity on attractant concentration in an exponential gradient. While maxima of the average drift velocity can be interpreted within analytical linear-response theory of chemotaxis in shallow gradients, limits in drift due to steep gradients and finite number of receptor-methylation sites for adaptation go beyond perturbation theory. For instance, we found a surprising pinning of the cells to the concentration in the gradient at which cells run out of methylation sites. 
To validate the positions of maximal drift, we recorded single-cell trajectories in carefully designed chemical gradients using microfluidics. 
\end{abstract}

\maketitle

\begin{figure}[ht]
\includegraphics[width=8.3cm]{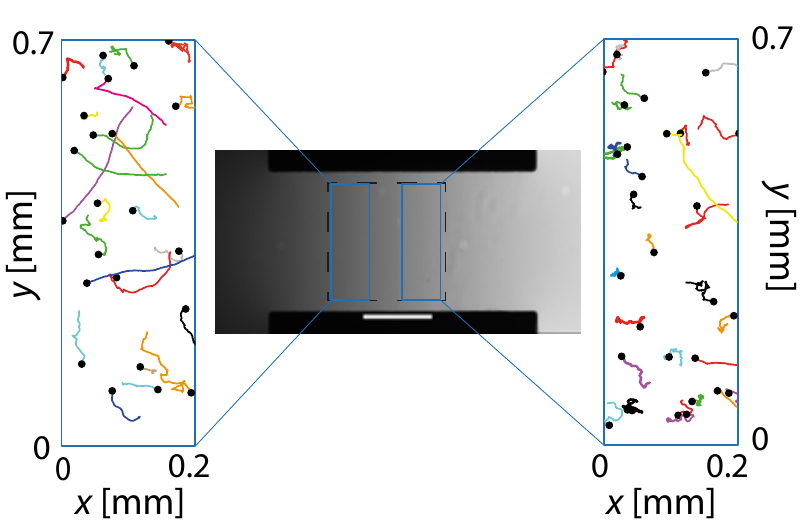}
\caption{{\bf Schematic of the experimental setup.}
A chemical gradient of $\alpha$-D,L-methylaspartic acid (MeAsp, a non-metabolizable analogue of the amino acid Asp) is created in a microfluidic device by maintaining a fixed concentration on one side of the channel and zero on the other. \textit{E. coli} cells (strain MG1655) are injected on both sides and free to move in aerobic conditions. The gradient is stable after about 1h 30min and the data were acquired after 2h, 3h and 4h with each experiment repeated 3 times \cite{SItext}. The gradient was measured after the final acquisition using fluorescein. (Middle) Fluorescence picture of the microfluidic chamber with the white bar representing $500$ $\mu$m. (Left and right) Exemplar of single-\textit{E. coli} trajectories from a typical movie, acquired in the middle of the channel with trajectory starting points marked with black dots (dashed box, for details see \cite{SItext}). Some of them are relatively straight (left) while others are curled (right). The MeAsp gradient is oriented to the right in this image (lighter shading corresponds to higher ligand concentration). The average concentration in the channel was 1mM. 
}
\label{fig1new}
\end{figure}

Cell behavior is notoriously difficult to interpret due to short observation times, variability, and dependence on experimental conditions. Take for instance the bacterium {\it E. coli}, which is able to swim up gradients of nutrients in a process called chemotaxis. Its swimming behavior is a result of sensing by cooperative mixed-receptor clusters, signaling by phosphorylation of a response regulator, adaptation by covalent receptor methylation, and motility by flagellated rotary motors \cite{SourjikWingreenRev}, operating on wide-ranging time scales. This bacterium's chemotaxis pathway has been extremely well characterized experimentally, but when conducting single-cell experiments using microfluidics in a simple linear chemical gradient of chemoattractant $\alpha$-D,L-methylaspartic acid (MeAsp), the obtained trajectories depict a complex structure in space and time (Fig. \ref{fig1new}, see caption for details). In some regions, trajectories are curled up due to quickly alternating periods of runs and random tumbles, while other regions show elongated trajectories due to more efficient chemotaxis up the gradient. How can this fine structure be understood quantitatively? 
Although phenotypic variability can explain different behaviors \cite{EmonetWaite16MSB}, here we demonstrate that even a single phenotype can show a range of unexpected behaviors.

One primary way to quantify the effectiveness of chemotaxis up a gradient is the determination of the drift velocity, defined as the cell's velocity component in the direction of the gradient. 
However, analytical calculations of average drift are generally hampered due to cellular memory from adaptation, limiting theoretical approaches to shallow gradients. 
As a result, previous approaches linearize pathways either around the adapted state or the steady state in exponential gradients when receptors are sensitive \cite{Reneaux_10, CelVerg2010, Tu10, Tu12, Dufour14, Colin14}. While {\it E. coli} is known to chemotax best in exponential gradients due to logarithmic dependence of the receptor free-energy on ligand concentration \cite{KalTuWu09, ShimTuBerg10, Micali14}, most previous approaches do not allow the prediction of spatially resolved drift, including outside the receptor sensitive region (for exceptions see \cite{Vergassola_PCB16, EmonetLong17PlosCB}).
Experimentally, drift is generally averaged over the whole population (across the whole observation chamber), removing any spatial dependence \cite{MassonCelaniVerg2012, Colin14}. 
Furthermore, since experimentally realized gradients are generally linear, there has been little comparison with theory in exponential gradients. 

In this {\it letter}, we studied chemotactic behavior in terms of the drift up the gradient. Using simulations in arbitrarily steep exponential gradients, we observed hard limits and a surprising level of spatial variation of the drift velocity, even for a single phenotype (a cell with a specific set of parameters). 
While drift in shallow gradients can be well explained by analytical theory, drift in steep gradients cannot, requiring the inclusion of nonlinear effects (receptor saturation, receptor insensitivity, imprecise adaptation).    
To validate some of our findings (positions of maximal drift), we tracked individual cells in a microfluidic device using carefully crafted linear gradients to match our exponential gradient from simulations \cite{SItext}. 

\begin{figure}[ht]
\includegraphics[width=8.3cm]{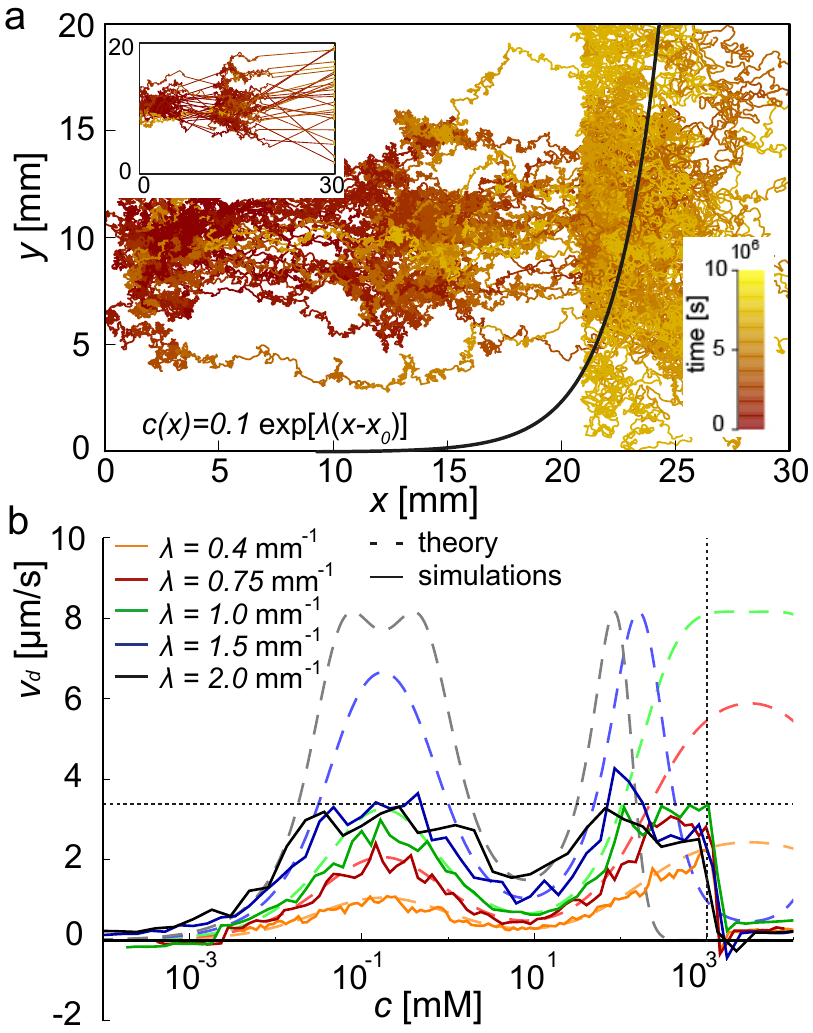}
\caption{{\bf Swimming behavior of {\it E. coli} cells from simulations.}
(a) Trajectories of cells chemotacting in an exponential gradient of MeAsp from our simulations based on a modified version of RapidCell software (see \cite{SItext} for details) \cite{VlaLovSou08}. Cells are initially placed at $x_0=2.5$ mm and $y_0=10$ mm (note that the \textit{y}-position is irrelevant as the gradient is independent of \textit{y}). We impose reflecting boundary conditions at $x=0$, $y=0$ and $y=20$ mm, and adsorbing boundary condition at $x=30$mm.
(inset) Trajectories without rotational diffusion. Relative gradient is $\lambda=0.75 \ \text{mm}^{-1}$. 
(b) Average drift velocity from simulations (solid lines) and analytical (dashed lines) theory.
Vertical and horizontal dotted lines indicate the discrepancy between theory and simulations due to finite number of methylation sites (vertical) and very steep gradients (horizontal). Note that this averaged single-cell drift from trajectories is equivalent to the steady-state drift from experiments (next figure). For both panels, parameters are reported in Table S1. 
}
\label{fig1}
\end{figure}

To gain insights into the fine structure of swimming bacteria in absence of phenotypic cell-to-cell variability, we conducted extensive simulations of \textit{E. coli} cells in exponential gradients of MeAsp using a modified version of RapidCell software (see \cite{SItext} for details) \cite{VlaLovSou08}. 
Specifically, we used exponential gradients $c(x)\sim\exp(-\lambda x)$ with constant relative gradient $\lambda=c^{-1} dc/dx$. 
To achieve this, we modeled the \textit{E. coli} chemotaxis pathway as follows: the receptor free-energy is given by the cooperative two-state receptors, modeled by the Monod-Wyman-Changeux model in which ligand binding inhibits receptor activity \cite{BrayLevMFirth98,DukeBray99,MelTu05,SouBerg04,KeyEndSko06}, the receptors methylation level evolves by integral feedback control for precise adaptation \cite{BarLei97, ClauOleEnd10}, the phosphorylated response regulator $\text{CheY}_\text{p}$ follows the receptor activity \cite{VlaLovSou08}, and the motor switches according to the ultrasensitive response measured in \cite{YuanBerg13}. Furthermore, the bacterial state is determined following observations in \cite{Chemla14Elife}, namely one motor rotating CW is enough to trigger a tumble and the number of motor is set to three (see \cite{SItext} for details of the pathway).  
The simulations further neglect interactions between cells and noise (in the extracellular concentration and in signaling), and assume a constant velocity during runs (see \cite{SItext} for further details); randomness is the result of stochastic motor switching, stochastically chosen tumble angles and stochastic reorientation due to rotational diffusion. 
We subsequently calculated the drift from long individual simulated trajectories (Fig. \ref{fig1}a) by taking the velocity component in the direction of the gradient at each position along the gradient. We averaged the drift velocity along the perpendicular direction ($y$) and plotted it as a function of the concentration sampled by the cells (which varies only in the direction along the gradient $x$). We observed two characteristic peaks (maxima, or plateaus in very steep gradients)  of the drift velocity (Fig. \ref{fig1}b), where the first (second) peak is due to Tar (Tsr) responding to MeAsp. Note that the drift velocity defined here is a property of individual trajectories while other common measures of the swimming behavior such as the chemotactic migration coefficient describe average population movement \cite{KalTuWu09,LazAhmShi2011}. Averaging the drift velocity of trajectories at a given location, and hence at a given concentration, allows us quantify for spatial heterogeneity of the trajectories. 

To explain these observations at the behavioral level, we first used common linear-response theory for the average drift to shallow gradients of arbitrary ligand concentrations (even outside the receptor sensitive regime). 
Similar to previous works  \cite{Dufour14, Colin14}, we linearized the pathway responses assuming small changes of receptor activity on a time scale $\delta t$ much shorter than the adaptation time ($\sim 10-20 s$). Note that adaptation is considered perfect for all concentrations in our analytical theory (but this assumption can also be relaxed \cite{Vergassola_PCB16}). 
Under these assumptions, the average drift velocity is given by
\begin{equation}
v_d=K(\left<A\right>) \frac{\partial F}{\partial c}\nabla c,\label{vd}
\end{equation}
where $\left<A\right>$ is the local average receptor activity (see Eq. S20 and Fig. S3 in \cite{SItext}). Briefly, in shallow gradients the average activity is the adapted activity, while in steep gradients it is further reduced. The susceptibility $K(\left<A\right>)$ includes the biological parameters (rate constants, adaptation time, sensitivity of receptors, and rotational diffusion - see \cite{SItext}). 
Symbol $\nabla c$ denotes the (local) gradient steepness. Importantly, $F$ is the receptor free-energy difference between the active and inactive receptor conformations \cite{KeyEndSko06}, leading to sensitivity $\partial F/\partial c=N \sum_{i=\{\text{Tar, Tsr}\} } \nu_i \frac{K^\textit{on}_i-K^\textit{off}_i}{(c+K^\textit{off}_i)(c+K^\textit{on}_i)}$ with $N$ the total number of receptors in a cluster, $\nu_i$ the fraction of receptor type $i$ for the most abundant receptor types Tar and Tsr, and $K^\text{on}_i$ and  $K^\text{off}_i$ the dissociation constants in the receptor \textit{on} and \textit{off} states. 
Product $\partial F/ \partial c \cdot \nabla c$ for exponential (not for linear) gradients exhibits the expected two peaks due to the sensitivity of the high and low affinity Tar and Tsr receptors, respectively (see \cite{SItext} and Fig. S1) \cite{SouBerg02a,EndWin06}. Unlike some previous theories \cite{Colin14}, our activity changes due to drift \cite{SItext,Tu12, Dufour14}.

As expected, for exponential gradients with small relative gradient $\lambda$, our analytical theory agrees well with simulations (Fig. \ref{fig1}b). However, for larger $\lambda$ significant differences appear. This is expected since our theory assumes only small variations of receptor activity on a time scale smaller than the adaptation time. In steep gradients, large changes in concentration result in large changes in the receptor activity and methylation level, invalidating our assumption of memory-less dynamics. 
In particular, we noticed that the drift from simulations never becomes larger than a threshold value of about $3.3 \ \mu\text{m/s} \approx v_0/3$ for the parameters from Fig. \ref{fig1}b, with $v_0=11.81 \ \mu$m/s the average run velocity extracted from experiments.  
This is because very steep gradients (large $\lambda$) lead to complete receptor inhibition and non-stop running, where rotational diffusion brings cells off course (see Eq. S23 in \cite{SItext}). 
Tumbles are required to restart chemotaxis, thus limiting drift.  
Faster adaptation rates increase this threshold value (see \cite{SItext} for effects of rotational diffusion and adaptation rates on drift) \cite{VlaLovSou08}. 

We also noticed that the drift declines sharply beyond a concentration of about $10^3$ mM (Fig. \ref{fig1}b and \cite{SItext}), which is due to the finite number of methylation sites available on a receptor. In our analytical theory the number of methylation sites is assumed to be infinite, while in simulations it is 8 in line with data \cite{SourjikBerg2002b}, leading to loss of adaptation beyond $\simeq 10^3$ mM \cite{Neumann14}. 
This concentration represents the semi-permeable boundary at $x\simeq22$ mm shown in Fig. \ref{fig1}a at around : cells coming from lower concentrations can move to higher concentrations, but then enter a state of non-stop running. Due to rotational diffusion these cells return eventually to concentrations of about $10^3$ mM, where they regain their sensitivity and eventually move up the gradient again. Indeed, once rotational diffusion is removed from the simulations, this semi-permeable boundary disappears (Fig. \ref{fig1}a, inset).

\begin{figure}[t!]
\includegraphics[width=8.3cm]{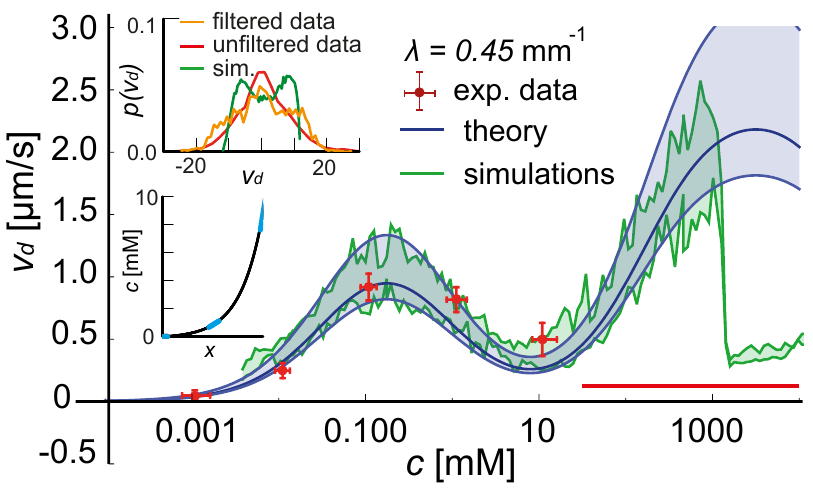}
\caption{{\bf Experimental verification of peak using microfluidics.}  
Comparison of drift from theory, simulations and experiments. Experimental data in linear gradients (red dots); each data point is the average over 9 measurements (3 measurements at different times in triplicates). Vertical bars are error of the mean, while horizontal bars range from low to high concentrations observed during experiments (relative gradients are about 0.45 mm$^{-1}$, cells swim with an average velocity of about $11.8 \pm 1.3 \ \mu$m s$^{-1}$ \cite{SItext}). Red bar indicates osmotic stress regime \cite{LazAhmShi2011}. Theory (blue area and lines) and simulations (green area and lines) in exponential gradient ($\lambda=0.45$ mm$^{-1}$) with upper ($13.1 \ \mu$m s$^{-1}$) and lower ($10.5 \ \mu$m s$^{-1}$) velocity bounds as shown \cite{SItext}. Parameters for theory and simulations are reported in Table S1. (Insets) Comparison of distributions of drift velocities [$\mu$m/s] from unfiltered (red line) and filtered (orange line) experimental trajectories, where the filter selected trajectories with average run velocity of $11.81 \pm 1.31 \ \mu$m/s, as well as trajectories from simulations (green line) for concentrations around 0.1 mM (top). The bimodal distribution from simulations is due to projection along the $x$-axis, while the center peak of the unfiltered data is due to tumbles of finite duration, with the peak height reduced when filtered for speed. (Bottom) Piecewise linear approximation (blue) of the exponential gradient used in the experiments.}
\label{fig2}
\end{figure}

Up to this date, most experimentally determined drift velocities were based on averaging over whole fields of cells, removing any spatial structure in cell behavior \cite{Colin14, EmonetLong17PlosCB} (see \cite{EmonetWaite16MSB} for an exception). In order to test the existence of the peaks, we used microfluidics (Fig. \ref{fig1new}; see caption and \cite{SItext} for details). As gradients are generally linear (and relatively shallow) in such devices, we designed linear gradients to match the relative gradient of the exponential gradient used in simulations (for moderate relative steepness $\lambda=0.45$ mM/mm). Specifically, we focused on ligand concentrations suitable for sampling the first peak and surrounding concentrations (see \cite{SItext} for details; note that larger MeAsp concentrations are difficult to apply due to toxicity from high osmolarity as observed before \cite{LazAhmShi2011}). Indeed, our experiments verify the Tar peak at the expected location (Fig. \ref{fig2}). Note this peak is much wider in \cite{KalTuWu09}, presumably due to spatial averaging using the chemotactic migration coefficient. 
Taken together our integrated approach of analytical theory, computer simulations and carefully designed microfluidics experiments aided the uncovering of bacterial cell behavior in exponential gradients. 

Even in the well-characterized bacterium \textit{E. coli}, behavior is remarkably complex. In this \textit{letter}, we explained the peaks and limits of the drift velocity underlying cell behavior. In the future, it would be interesting to investigate how regions in a gradient of high drift fundamentally emerge from information gain and energy consumption \cite{Micali16,TuLan16Review}. Up to now, this issue is either investigated heuristically (see Fig. S8 in \cite{SItext}) or with very simple models \cite{DePalo13, Barato2014, tenWolde14PNAS}.

\ \\
\noindent{\bf Author Contributions:} Conceived and designed the experiments: GM RC VS RGE. Performed the experiments: RC. Analyzed the data: GM RC. Contributed reagents/materials/analysis tools: GM RC VS RGE. Wrote the paper: GM RC VS RGE. 
\ \\

\ \\
\noindent{\bf Acknowledgements:} GM and RGE thankfully acknowledge support from the European Research Council Starting Grant N. 280492-PPHPI. R.C. and V.S. acknowledge support from the European Research Council Advanced Grant 294761-MicRobE. We thank M. Koch, K. Volz and W. Stolz for providing access to the microfabrication facility. 
\ \\

\setcounter{equation}{0}
\renewcommand\theequation{S\arabic{equation}}
\setcounter{figure}{0}
\renewcommand\thefigure{S\arabic{figure}}
\setcounter{table}{0}
\renewcommand\thetable{S\arabic{table}}

\section*{SUPPLEMENTARY MATERIAL}

\section{Basic model for \textit{E. coli} chemotaxis} 

\noindent \textit{E. coli} moves in a run-and-tumble motion using a combination of straight swims affected by rotational diffusion (run) and random reorientations (tumbles). Chemotaxis is the ability of the bacteria to bias their otherwise random walk in the direction of the chemical attractant. Here, we briefly review the model for \textit{E. coli} chemotaxis (additional details can be found in recent reviews such as \cite{SourjikWingreenRev, Micali16}). The extracellular ligand molecules bind and unbind the receptors, which can be in either active or inactive conformations. The active receptors trigger the auto-phosphorylation of the internal protein CheA attached to the receptors. $\text{CheA}_\text{p}$ in turn phosphorylates the response regulator cytoplasmic protein CheY. $\text{CheY}_\text{p}$ regulates the motor rotation by binding to the internal part of the motor by promoting a change in motor rotation from default counterclockwise (CCW) to clockwise (CW) rotation. 
The rotation of the 5-8 motors determinates the run and tumble state of the cell. 
The receptor activity adapts perfectly to constant stimulation of attractant aspartate and non metabolizable attractant $\alpha$-D,L-methylaspartic acid (MeAsp), i.e. the activity returns to the value before stimulations (adapted value $A^*$). This is achieved by methylation of inactive and de-metylation of active receptors by enzymes CheR and $\text{CheB}_\text{p}$, respectively. In particular, CheB phosphorylation through $\text{CheA}_\text{p}$, and this integral feedback control guarantees perfect adaptation for aspartate and MeAsp but not for other attractants, e.g. serine \cite{BarLei97, Neumann14}. 

\ \\

\noindent A well established model for the overall receptor activity of chemoreceptors is the Monod-Wyman-Changeux (MWC) model \cite{MWC65, KeyEndSko06}. Following the MWC model, the receptor activity is given by
\begin{align}
A(m,c) &=\left[1+e^{F(m,c)}\right]^{-1}, 
\label{Eq:A}
\end{align}
where $m$ is the methylation level of the receptors, $c$ is the concentration of ligand, and $F(m,c)$ is the receptor free-energy difference, given by
\begin{align}
F(m,c) &= N \left[ \epsilon(m) +\sum_{i \in \{ \text{Tar,Tsr} \} } \nu_i \log \frac{1+ c/K^\text{off}_i }{1+c/K^\text{on}_i} \right],
\end{align}
where $\epsilon(m)=1-\gamma m$, with $\gamma$ a constant, $N$ is the number of receptors in a cluster, $\nu_i$ is the fraction of Tar and Tsr receptors, and $K^{\text{on}}$ and $K^{\text{off}}$ are the dissociation constants for the active and inactive states of the receptors, respectively \cite{KeyEndSko06}. 
The methylation dynamics, following integral feedback control, are given by \cite{BarLei97, ClauOleEnd10}
\begin{align}
\label{Eq:m}
\frac{\text{d}m}{\text{d}t} = \gamma_R \left[ 1-A(t) \right] - \gamma_B A(t)^3, 
\end{align}
where $\gamma_R$ and $\gamma_B$ are the methylation and de-methylation rates, respectively. 
The phosphorylation of $\text{CheY}$ happens fast and thus the $\text{CheY}_\text{p}$ concentration is normally assumed to be at quasi-steady state \cite{VlaLovSou08}, given by 
\begin{align}
\label{Eq:Yp}
Y^p=\frac{k_Y A Y^T}{k_Y A+k_Z Z + \gamma_Y},
\end{align}
where $Y^T$ and $Z$ are the total concentration of CheY and CheZ, respectively, and $k_Y$, $k_Z$, and $\gamma_Y$ are the phosphorylation, de-phosphorylation and degradation rates of CheY, respectively. The rate of switching rotation of one single motor is given by 
\begin{align}
\label{Eq:kccwTOcw}
k_{\text{CCW} \rightarrow \text{CW}} &= \frac{1}{t_0} e^{ - \left(20-40\frac{Y^p}{Y^p+3.06}\right)} \\
\label{Eq:kcwTOccw}
k_{\text{CW} \rightarrow \text{CCW}} &= \frac{1}{t_0} e^{+ \left(20-40\frac{Y^p}{Y^p+3.06}\right)}, 
\end{align}
where the $t^{-1}_0=1.3 \ \text{s}^{-1}$ is the switching frequency of the motor (modeled as a bistable system) in order to match the experimental motor switching rates \cite{CluSurLei00, Sneddon12, YuanBerg13, Dufour14}. The most recent model states that the \textit{E. coli} motility can be explained by a `veto' model over an effective number of $M$ motors (around 3), in which one receptor rotating CW is enough to trigger a tumble \cite{Chemla14}. Thus, the rate of switching from run to tumble state is 
\begin{align}
\label{Eq:krt}
k_{\text{r} \rightarrow \text{t}} &= (d-1) D_r +  \frac{1}{t_r}  M e^{ -\left(20-40\frac{Y^p}{Y^p+3.06}\right)},
\end{align}
where the term $(d-1) D_r$ accounts for rotational diffusion ($d$ number of dimension). All motors need to rotate CCW to trigger a run. However assuming that most of the time only one motor would be rotating CW during a tumble, the rate of switching from a tumble to a run is 
\begin{align}
\label{Eq:ktr}
k_{\text{t} \rightarrow \text{r}} &=  \frac{1}{t_r}  e^{ \left(20-40\frac{Y^p}{Y^p+3.06}\right)},
\end{align}
with $t_r=t_0 (1 + \alpha/d)$ to include a non-uniform angle distribution after a tumble \cite{CelVerg2010} for $\alpha=0.37$ and $d=2$ dimensions, matching the average experimentally measured angle after a tumble \cite{berg1972chemotaxis, VlaLovSou08}. To explain the connection between parameter $\alpha$ and the angle distribution more note that the adapted run time $t_0$ is a parameter in both our analytical theory and simulations. However, in our simulations there is a preference for cells to move in a similar directions after a tumble as before a tumble, especially if only one or two motors switch to CW rotation. This directional bias is however not part of the analytical model. To compensate for this effect, parameter $\alpha$ is introduced to effectively increase the run time to reflect this directional bias ($t_r > t_0$).

\section{Analytical theory for the drift velocity}

\noindent Here, we present the details of calculations that lead to Eq. (1) in the main text. The calculations were performed on paper and checked with MATHEMATICA 8. The theory for the drift velocity of a cell swimming in a given concentration $c$ and gradient $\nabla c$ of MeAsp is based on the linearization of the \textit{E. coli} pathway. Note that the theory assumes perfect adaptation, which is a good approximation for MeAsp but not for other attractants, and an infinite number of receptor methylation sites.  

\ \\

\noindent In our model for the \textit{E. coli} chemotaxis pathway, the receptor activity, $A$, given by Eq. \eqref{Eq:A} is a function of the receptor free-energy difference, $F$, which in turn depends on methylation level $m$ and attractant concentration $c$. Thus, linearizing the activity around the initial value $A_0$ at position $(x_0,y_0)$ for small changes in time provides
\begin{align}
\delta A = \left. \frac{\partial A}{\partial F} \right|_0 \left( \left. \frac{\partial F}{\partial m}\right|_0 \delta m + \left.\frac{\partial F}{\partial c}\right|_0 \delta c \right),
\label{Eq:dA}
\end{align}
where 
\begin{align*}
\left. \frac{\partial A}{\partial F} \right|_0 &= -  \frac{ e^{F_0}}{\left(1+e^{F_0} \right)^2} = A_0 (1-A_0), \\
\left. \frac{\partial F}{\partial m}\right|_0 &= N \frac{\text{d} \epsilon}{\text{d}m} = - N \gamma. \\ 
\end{align*}
The change in concentration $\delta c$ can be easily expressed as a change in time
\begin{align}
\delta c = \left. \nabla c \right|_0 \cdot \bar{v}_0 \ \delta t, 
\label{Eq:dc}
\end{align}
where $\bar{v}_0$ is the vector representing the instantaneous run velocity of the cell, which is assumed to have constant speed for all the runs. 
The change of methylation level in $\delta t$ is calculated from the equation for the methylation and de-methylation dynamics, Eq. \eqref{Eq:m}. 
Thus, $\delta m$ is obtained by solving 
\begin{align}
\frac{ \text{d} (\delta m)}{\text{d}t} & = -\left[\gamma_R + 3 A_0^2 \gamma_B \right] \delta A = - \zeta \delta A,
\label{Eq:dm}
\end{align} 
where $\zeta:=(\gamma_R + 3 A_0^2 \gamma_B)$.  
Inserting Eqs. \eqref{Eq:dA} and \eqref{Eq:dc} in Eq. \eqref{Eq:dm} gives
\begin{align}
\nonumber
\frac{ \text{d} (\delta m)}{\text{d}t} & = - \zeta  \left. \frac{\partial A}{\partial F} \right|_0 \left( \left. \frac{\partial F}{\partial m} \right|_0 \delta m +  \left. \frac{\partial F}{\partial c} \right|_0 \left. \nabla c \right|_0 \cdot \bar{v}_0 \delta t \right) \\ 
\nonumber
& = - \zeta \left( 1-A_0 \right) A_0 \left( N \gamma \delta m - \left. \frac{\partial F}{\partial c} \right|_0 \left. \nabla c \right|_0 \cdot \bar{v}_0 \delta t \right).
\end{align}
This equation can be solved and provides an expression for $\delta m$,
\begin{align}
\delta m & =  \frac{ \left. \frac{\partial F}{\partial c} \right|_0 \left. \cdot \nabla c \right|_0 \cdot \bar{v}_0  }{N \gamma}   \left( \delta t -\tau_m(A_0)+ \tau_m(A_0) e^{-\frac{ \delta t}{\tau_m(A_0)}} \right),
\label{Eq:dmF}
\end{align}
where the adaptation time is defined by $\tau_m(A_0):=\left(\zeta A_0 (1-A_0) N \gamma\right)^{-1}$.  
The variation of $\text{CheY}_\text{p}$ ($Y^p$) is given by linearizing Eq. \eqref{Eq:Yp}, and thus 
\begin{align}
\nonumber
\delta Y^p &= \left. \frac{\partial Y^p}{\partial A} \right|_0 \delta A = \\
\nonumber
& =  \left. \frac{\partial Y^p}{\partial A} \right|_0 (1-A_0) A_0  \tau_m(A_0) \left. \frac{\partial F}{\partial c} \right|_0 \left. \nabla c \right|_0  \cdot \bar{v}_0   
\left(  1 -  e^{-\frac{ \delta t}{\tau_m(A_0)}}  \right) = \\ 
&= \frac{ k_Y (\gamma_Y + k_Z Z)}{(\gamma_Y + A_0 k_Y + k_Z Z)^2} Y^T  (1-A_0) A_0 \tau_m(A_0)  \left. \frac{\partial F}{\partial c} \right|_0 \left. \nabla c \right|_0 \cdot \bar{v}_0   \left(  1 -  e^{-\frac{ \delta t}{\tau_m(A_0)}}  \right), 
\label{Eq:dY}
\end{align}
where Eqs. \eqref{Eq:dc} and \eqref{Eq:dmF} are used in Eq. \eqref{Eq:dA}. 
For simplicity, Eq. \eqref{Eq:dY} can be written 
\[
\delta Y^p = K'(A_0) \left(  1 -  e^{-\frac{ \delta t}{\tau_m(A_0)}}  \right),
\]
where $K'$ is defined by dividing the right-hand side of Eq. \eqref{Eq:dY} by $\left(1 -  e^{-\frac{ \delta t}{\tau_m(A_0)}}  \right)$.
The change of rate $k_{\text{CCW} \rightarrow \text{CW}}$ ($k_{\text{CW} \rightarrow \text{CCW}}$) from Eq. \eqref{Eq:kccwTOcw} (Eq. \eqref{Eq:kcwTOccw}) in a small time $\delta t$ is given by
\begin{align}
\delta k_{\text{CCW} \rightarrow \text{CW}} & = \left. \frac{\partial k_{\text{CCW} \rightarrow \text{CW}}}{\partial Y^p} \right|_0 \delta Y^p =  \left. \frac{\partial k_{\text{CCW} \rightarrow \text{CW}}}{\partial Y^p} \right|_0 K'(A_0) \left(  1 -  e^{-\frac{ \delta t}{\tau_m(A_0)}}  \right) 
\end{align}
(with a similar expression for $\delta k_{\text{CW} \rightarrow \text{CCW}}$). 
The total time of a run is given by
\begin{align}
\tau_r = \int_0^{\infty} \text{d}t \ e^{-\int_0^t \text{d}t' \ k_{\text{r} \rightarrow \text{t}} }
\label{Eq:taur}
\end{align}
(and similarly, the total time of a tumble, $\tau_t$, is given by using $k_{\text{t} \rightarrow \text{r}}$ instead). 
In the linear regime of CheY$_\text{p}$ changes, only the motor response changes, while the rotational diffusion and the angle after a tumble do not depend on CheY$_\text{p}$. Thus, the integral over $t'$ gives
\begin{align*}
\int_0^t \text{d}t' \ k_{\text{r} \rightarrow \text{t}} &= \left. k_{\text{r} \rightarrow \text{t}} \right|_0 t + \left. \frac{\partial k_{\text{CCW} \rightarrow \text{CW}}}{\partial Y^p} \right|_0 K'(A_0) \left( t - \tau_m(A_0) + \tau_m(A_0) e^{-\frac{t}{\tau_m(A_0)}}\right).  
\end{align*} 
Thus, Eq. \eqref{Eq:taur} becomes
\begin{align}
\nonumber
\tau_r &\simeq \int_0^{\infty} \text{d}t \ e^{-\left. k_{\text{r} \rightarrow \text{t}} \right|_0 t} \left[1 +  \left. \frac{\partial k_{\text{CCW} \rightarrow \text{CW}}}{\partial Y^p} \right|_0 K'(A_0) \left( t - \tau_m(A_0) + \tau_m(A_0) e^{-\frac{t}{\tau_m(A_0)}}\right)  \right] \\ 
\nonumber
& =  \frac{1}{\left. k_{\text{r} \rightarrow \text{t}} \right|_0} + \frac{ \left. \frac{\partial k_{\text{CCW} \rightarrow \text{CW}}}{\partial Y^p} \right|_0 K'(A_0)}{ \left. k_{\text{r} \rightarrow \text{t}} \right|_0 \left( 1+ \tau_m(A_0) \left. k_{\text{r} \rightarrow \text{t}} \right|_0 \right)} \\ 
& = \frac{1}{\left. k_{\text{r} \rightarrow \text{t}} \right|_0} + \frac{K''(A_0) \left. \nabla c \right|_0 \cdot \bar{v}_0 }{ \left. k_{\text{r} \rightarrow \text{t}} \right|_0 \left( 1+ \tau_m(A_0) \left. k_{\text{r} \rightarrow \text{t}} \right|_0 \right)},
\end{align}
valid in the linear regime only, i.e. $\left. (\delta A/A) \right|_0 \ll 1$, $\left. (\delta Y^p/Y^p) \right|_0 \ll 1$, $\left. (\delta k_{\text{CCW} \rightarrow \text{CW}}/k_{\text{CCW} \rightarrow \text{CW}}) \right|_0 \ll 1$ and $\left. (\delta k_{\text{CW} \rightarrow \text{CCW}}/k_{\text{CW} \rightarrow \text{CCW}}) \right|_0 \ll 1$.  
Note that  
\[
K''(A_0):= \left. \frac{\partial k_{\text{CCW} \rightarrow \text{CW}}}{\partial Y^p} \right|_0 \frac{ K'(A_0) }{\left. \nabla c \right|_0 \cdot \bar{v}_0 },
\]
and that $\left. \nabla c \right|_0 \cdot \bar{v}_0 = \left. \nabla c \right|_0 v_0 \cos \theta$, where $\theta$ is the angle between the gradient and the direction of running of the cell. 

\ \\

\begin{figure}[ht]
\includegraphics[width=\textwidth]{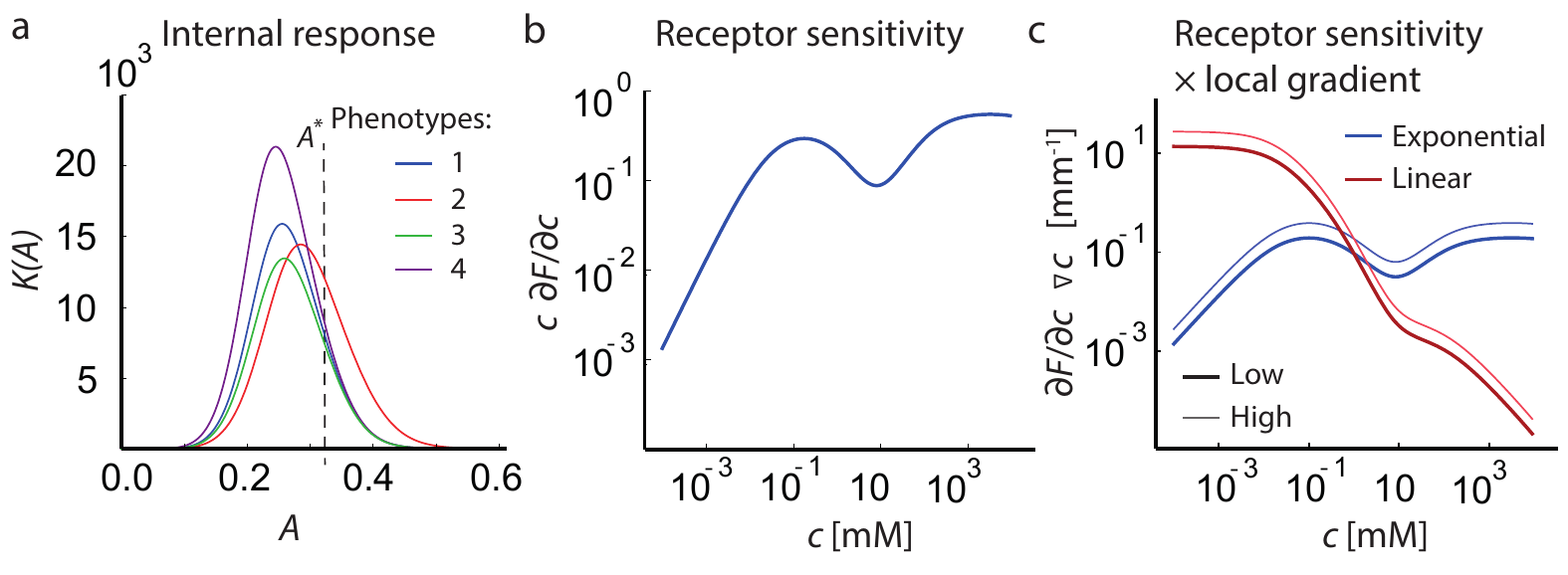}
\caption{{\bf Analytical theory for drift velocity.}
(a) Internal susceptibility, $K(A)$, as a function of the receptor activity, $A$. Each curve corresponds to a single phenotype (set by parameter values). Phenotype 1 (as in Fig. 1): $\text{CheY}^\text{tot}$= 7.9 $\mu$M, $D_r = 0.062$ rad$^2$/s, $\tau_m = 13.7$ s (blue); Phenotype 2: $\text{CheY}^\text{tot}$= 7.3 $\mu$M, $D_r$= 0.062 rad$^2$/s, $\tau_m$= 13.7 s (red); Phenotype 3: $\text{CheY}^\text{tot}$= 7.9 $\mu$M, $D_r$= 0.062 rad/s, $\tau_m$= 8.7 s (green); Phenotype 4: $\text{CheY}^\text{tot}$= 7.9 $\mu$M, $D_r$= 0.042 rad/s, $\tau_m$= 13.7 s (purple). The adapted activity, $A^*$, is not the activity which maximizes $K(A)$. (b) Methylation independent receptor sensitivity, $c \cdot \partial F / \partial c$, as a function of the external concentration. (c) Product of methylation-independent receptor sensitivity and local gradient, $\partial F/\partial c \cdot \nabla c$, for low (thick line) and high (thin line) exponential gradients (blue) and linear gradients (red) as a function of the external concentration $c$.
}
\label{fig1bis}
\end{figure}

\noindent The drift velocity in $2d$ is then given by 
\begin{align}
\label{Eq:vdpre}
v_d = \frac{ \int_{0}^{2\pi} \text{d}\theta \ \tau_r(\theta) \cos(\theta)}{\int_{0}^{2\pi} \text{d}\theta \ \tau_r(\theta) + \int_{0}^{2\pi} \text{d}\theta \ \tau_t(\theta) } v_0,
\end{align}
i.e. the ratio between the time spent running up the gradient minus the time spent running down the gradient, divided by the total time spent running and tumbling. 
In the numerator of Eq. \eqref{Eq:vdpre}, the zero-order terms, which are independent of $\theta$, vanish, and the the drift becomes 
\begin{align}
\left< \bar{v}_d \right>=K(A_0) \left. \frac{\partial F}{\partial c} \right|_0 \left. \nabla c \right|_0, 
\label{eq:driftN}
\end{align}
with the susceptibility 
\begin{align}
\nonumber
K(A_0) &= \frac{\left. k_{\text{t} \rightarrow \text{r}} \right|_0 K''(A_0) \ v_0}{d  \left. k_{\text{r} \rightarrow \text{t}} \right|_0 \left( 1 + \tau_m(A_0) \left. k_{\text{r} \rightarrow \text{t}} \right|_0 \right) \left( \left. k_{\text{r} \rightarrow \text{t}} \right|_0  + \left. k_{\text{t} \rightarrow \text{r}} \right|_0 \right) } \\
\label{Eq:KA}
&= \frac{ \left. k_{\text{t} \rightarrow \text{r}} \right|_0 \left. \frac{\partial k_{\text{CCW} \rightarrow \text{CW}}}{\partial Y^p} \right|_0  
\frac{ k_Y (\gamma_Y + k_Z Z)}{(\gamma_Y + A_0 k_Y + k_Z Z)^2} Y^T  (1-A_0) A_0 \tau_m(A_0) v_0^2}
{d  \left. k_{\text{r} \rightarrow \text{t}} \right|_0 \left( 1 + \tau_m(A_0) \left. k_{\text{r} \rightarrow \text{t}} \right|_0 \right) \left(\left. k_{\text{r} \rightarrow \text{t}} \right|_0  +\left. k_{\text{t} \rightarrow \text{r}} \right|_0 \right) } ,
\end{align} 
where $d$ is the number of dimensions (see \cite{Lovely75}). 
Note that in Eq. \eqref{eq:driftN}, 
\begin{equation}
\left. \frac{\partial F}{\partial c} \right|_0 \left. \nabla c \right|_0 = \left. c \frac{\partial F}{\partial c} \right|_0 \left. \frac{\nabla c}{c} \right|_0 = N \left[ \nu_a  \frac{ c \left(K_a^{\text{on}} - K_a^{\text{off}} \right)} { (c+K_a^{\text{off}}) (c+K_a^{\text{on}})} + \nu_s  \frac{ c \left( K_s^{\text{on}} - K_s^{\text{off}}\right) } { (c+K_s^{\text{off}}) (c+K_s^{\text{on}}) } 
 \right] \frac{\nabla c}{c}
\end{equation}
is the methylation-independent receptor sensitivity times the local relative gradient and thus it is a property of the gradient and receptors only, while $K(A_0)$ includes the memory dependence (Fig. \ref{fig1bis}). 

\ \\

\noindent The activity $A_0$ around which the drift velocity is calculated is normally the adapted activity $A^*$ \cite{Colin14}. However, the theory presented here is general and can apply to any $A_0$. Furthermore, knowing the drift velocity in the linear regime, the average activity $\left< A \right>$ can be written as
\begin{align}
\nonumber
\left< A \right> &= A_0 + \delta A \\
\label{Eq:Aav}
 & \simeq A_0 + (1-A_0) A_0  \tau_m(A_0) \left. \frac{\partial F}{\partial c} \right|_0 \left. \nabla c \right|_0  \left< \bar{v}_d \right>.   
 \end{align}
Valid in the linear regime only. The analytical drift velocity given in Eq. \eqref{eq:driftN} is then Eq. (1) of the main text. For shallow gradients $\left<A\right> \approx A_0 \approx A^*$, as often assumed in the literature \cite{Reneaux_10, CelVerg2010, Colin14}.

\begin{table}[h]
\begin{center}
\renewcommand*{\thempfootnote}{\fnsymbol{mpfootnote}}
   \begin{tabular}{| l l l l |}
       \hline
    & Symbol & value & reference \\ \hline
     {Cooperative receptor number} & $N$ & $13 \ (\in [5;16])$ & { \cite{EndSouWin2008}}  \\ 
     {Methylation free-energy} & $\epsilon(m)$ & $\in [-2.6;0.25]$ & { \cite{EndSouWin2008}}  \\ 
     {Fraction of Tar receptors} & $\nu_a$ & $\simeq 1/3$  & { \cite{KeyEndSko06}}  \\
     {Fraction of Tsr receptors} & $\nu_s$ & $\simeq 2/3$ & { \cite{KeyEndSko06}}  \\
     {Active receptors dissociation constant Tar\footnote[1]{All $K$ here are for Me-Asp}} & $K^\text{on}_a$ & 1.0 mM  &  { \cite{SouBerg04}}  \\ 
     {Active receptors dissociation constant Tsr} & $k^\text{on}_s$ &  $10^6$ mM &  { \cite{KeyEndSko06}}  \\ 
     {Inactive receptors dissociation constant Tar} & $k^\text{off}_a$ & 0.03 mM  &  { \cite{SouBerg04}}  \\ 
     { Inactive receptors dissociation constant Tsr} & $k^\text{off}_s$ & 100 mM  &  { \cite{KeyEndSko06}}  \\ 
     { Methylation rate}  & $g_R$ & $0.0069 \ s^{-1}$ & { \cite{ClauOleEnd10}} \\ 
     {De-methylation rate} & $g_B$ & $0.12 \ s^{-1}$ & { \cite{ClauOleEnd10}} \\ 
     {Phosphorylation rate of $Y_p$}  & $k_Y$ & 100 $\mu \text{M}^{-1} \text{s}^{-1}$ & { \cite{VlaLovSou08}} \\ 
     {De-phosphorylation rate of $Y_p$}  & $k_Z$ & 30 $Z^{-1} \text{s}^{-1}$ & { \cite{VlaLovSou08}} \\
     {Degradation rate of $Y_p$} & $\gamma_Y$ & 0.1 & { \cite{VlaLovSou08}} \\
     {Adapted activity} & $A^*$ & $\simeq 0.3$ & { \cite{VlaLovSou08}} \\ 
     {Total concentration of CheY} & $Y^T$ & $7.9 \ (\in [6;9.7]) \ \mu \text{M}$& { \cite{VlaLovSou08, Dufour14}} \\
     {Concentration of CheZ} & $Z$ & & { \cite{VlaLovSou08}}  \\ 
     {Motor Hill coefficient} & $m$ & $\in [15;22]$ & { \cite{YuanBerg13}} \\ 
     {Motor switching constant} & $t_0^{-1}$ & $1.3 \ \text{s}^{-1}$ & { \cite{Sneddon12}} \\ 
     {Corrections for average angle after a tumble} & $\alpha$ & $0.37 $ & {extrapolated from \cite{VlaLovSou08, CelVerg2010, Vergassola_PCB16}} \\ 
     {Motor dissociation constant} & $K$ & $\simeq 3 \ \mu \text{M}$ & { \cite{YuanBerg13}} \\ 
     {Rotational diffusion} & $D_r$ & $\simeq 0.062$ rad$^2$/s & { \cite{VlaLovSou08}} \\
     {Adaptation time} & $\tau_m$ &  $\tau_m(g_R, g_B, A^*)$ &  \\ 
     {Run velocity} & $v_0$ &  $\simeq 11.81 \text{[}\mu\text{m/\text{s}]}$ & { Fig. S4b}  \\ \hline
  \end{tabular}
 \end{center}
\caption{{\bf Parameters of \textit{E. coli} chemotaxis.} 
This table contains lists of maths symbols their values used and corresponding references.  
\label{tab:ChemPar}}
\end{table}

\section{Computational simulations}

\noindent 
The computational simulations are performed in Java and analysed in R. The Java code is based on \textit{RapidCell} 1.4.2 published in \cite{VlaLovSou08}. This software has the advantage of simulating large number of cells with relatively low computational time. This is achieved assuming the internal pathway to be at \textit{quasi}-steady state, thus avoiding to solve differential equations for the internal dynamics of CheY (as in Eq. \eqref{Eq:Yp}). Some adjustment to the code have been made following the model in Eqs. (S1-S6). The simulations do not account for the interactions between cells. 

\ \\

\noindent \textbf{Setting of the initial conditions.} There are several parameters that need to be set before starting simulations. 
\begin{itemize}
\item Simulations space. The simulations are normally produced inside a rectangular box of size set by us with periodic boundary conditions on the borders perpendicular to the gradient direction and adsorbing boundary conditions on the borders in the direction parallel to the gradient. 
\item Ligand concentration inside the box. The concentrations and gradients inside the box can be arbitrary and are set for the purpose of the simulation, $c(x,y)$, e.g. linear and exponential gradients, without noise (see \cite{Junhua16} for noise effects on drift). 
\item Biochemical parameters and rotational diffusion. All parameters appearing in Table \ref{tab:ChemPar} need to be set, including the receptor number and their sensitivity, the reaction rates for CheY phosphorylation and for the methylation/de-methylation dynamics, and the motor switching constant rates.
\item Cells parameters. The number of cells, the number of rotary motors, the run velocity (constant) and the initial state of each cell need to be specified. Normally, simulations involve from 30 to 100 cells, the run velocity $v_0$ is extracted from data (see later section), and each cell has three rotary motors. The state of each cell is specified by an array which includes the time ($t$), the position ($x$) and ($y$), the orientation ($\theta$), the concentration experienced by the cell ($c$), the receptor activity ($A$), the methylation level of the receptors ($m$), the CCW bias (CCWb), the number of motors moving CW (nCW) and the run or tumble state of the cell (RoT). All cells are initialized at the same position and in the same state, given by
\begin{align*} 
\Big( t_0& =0 \ \text{s}, \ x_0=1\ \text{cm}, \ y_0=2.5\ \text{mm}, \ \theta= \text{random} \in [0,2 \pi], \ c_0=c(x_0,y_0), \\
& \ A_0=A^*, \ m_0 \ : \ A(c_0,m_0)=A*, \ \text{CCWb}_0=\text{CCWb}^* , \ \text{nCW}_0 =0, \ \text{RoT}_0=\text{run} \Big), 
\end{align*}  
where the asterisk identifies the adapted quantities.
\end{itemize}

\ \\

\noindent \textbf{Description of the code for simulations.}  At each time step in the simulation, the state of the cells evolves as follow: 
\begin{itemize}
\item The run or tumble state of the cell is evaluated based on the state of the motors. We introduced the veto model in which one motor in the CW state is enough for a tumble. 
\item Based on the cell state, the position and the orientation are updated. If the RoT state is in a `tumble', the position does not change and the orientation $\theta$ is randomly assigned, $\theta \in [0,2 \pi]$ and the probability distribution depends on the number of motor rotating CW \cite{VlaLovSou08}. If the RoT state is in a `run', the position is updated: $x_t= x_{t-\delta t} + v_{r} \delta t \cos (\theta_{t-\delta t})$, $y_t= y_{t-\delta t} + v_{r} \delta t \sin( \theta_{t-\delta t})$. The orientation is then updated in order to account for rotational diffusion: $\theta_t = \theta_{t-\delta t} + \delta \theta$, where $\delta \theta$ is Gaussian distributed with mean given by the actual orientation and standard deviation $D_r$. 
\item The concentration experienced is updated given the new position in the box, $c_t = c(x_t, y_t)$.  
\item The receptor activity is updated given the concentration and the methylation level, following Eqs. (S1,S2), i.e. $A_t=\left\{1+ \exp \left[ N \left( \epsilon(m_{t-{\delta t}}) + \sum_{i \in \{ \text{Tar,Tsr} \} } \nu_i \log \frac{1+ c_{t}/k^\text{off}_i }{1+c_{t}/k^\text{on}_i} \right) \right] \right\}^{-1}$. 
\item The methylation level is updated following the ODE given in Eq. \eqref{Eq:m}, i.e. $m_t=m_{t-\delta t} + g_R (1-A_t) \delta t - g_B A_t^3 \delta t$.
\item The rate of switching from CCW to CW and from CW to CCW are updated following Eq. \eqref{Eq:kccwTOcw} and \eqref{Eq:kcwTOccw}, respectively, i.e.  $k^t_{\text{CCW}\rightarrow \text{CW}} = 1.3 \exp \left\{ -\left(20-40\frac{Y_t^p}{Y_t^p+3.06}\right) \right\}$ and $k^t_{\text{CW}\rightarrow \text{CCW}} = 1.3 \exp \left\{ +\left(20-40\frac{Y_t^p}{Y_t^p+3.06}\right) \right\}$, where  $Y^p_t=\frac{k_Y A_t Y^T}{k_Y A_t+k_Z Z + \gamma_Y}$. 
Hence, $\text{CCWb}_t= \frac{(Y^p_t)^m}{(K)^m+(Y^p_t)^m}$. 
\item The motor state is updated based on the probability given by the rate of switching multiplied by the time step. 
\item The time is updated $t=t+\delta t$.
\end{itemize}
For the definitions of the rates and parameters involved see Table \ref{tab:ChemPar}. 

\ \\

\noindent \textbf{Simulation output file.} The simulation produces an output file in which for each cell, at each time, a row represents the state of the cell, normally given by $(\text{cell number}, t, x, y,\theta,$ $c, A, m, \text{CCWb}, \text{nCW}, \text{RoT})$ or by a subset of interest. However, the output does not include the state for each time of the simulation, to avoid large files. Normally, the simulation time step $\delta t=0.001$ s and the output lines are produced each 100 time steps. Note, however, that for the evaluation of the fluctuation theorem, the output is produced at each time step (see last section).

\ \\

\noindent \textbf{Evaluation of the drift velocity.} The output file is analyzed using R. The box is divided in $n$ bins perpendicular to the gradient after a full simulation. The trajectories inside a bin are considered as different trajectories and the drift velocity is calculated over each of these (pieces of) trajectories. For the $i$-th pieces of a trajectory we compare the start and end points and calculate the local drift velocity as following 
\begin{align*}
v_{d,i} &= \frac{\Delta x_i}{\Delta t},
\end{align*}
where $x$ is the direction of the gradient and $\Delta$ denotes the difference between the end and start points. The average drift velocity is then calculated by averaging over all the trajectories inside a bin at the location of interest. 

\ \\

\begin{figure}[h!]
\includegraphics[width=\textwidth]{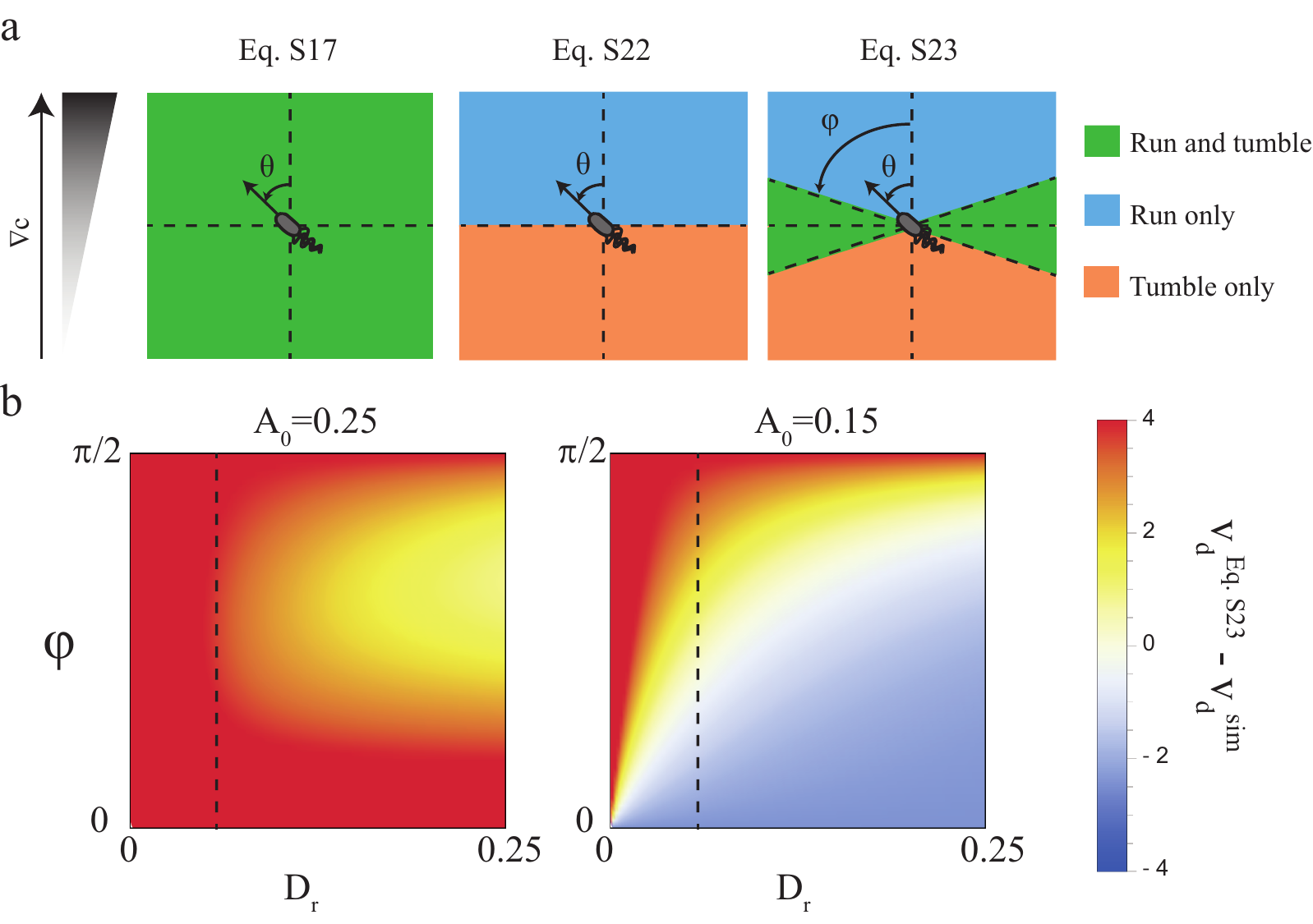}
\caption{{\bf Limits of drift velocity.} (a) The three equations proposed in this section provide different levels of approximation to the limit of the drift velocity shown in Fig. 2b of the main text. Eq. \eqref{Eq:vdpre} is derived from linear-response theory and applies to chemotactic cells that are fully sensitive over all the possible run directions (left). Eq. \eqref{eq:driftSat1} applies to insensitive cells which only run up the gradient (although their direction is affected rotational diffusion), and only tumble if their direction is down the gradient (middle). Eq. \eqref{eq:driftSat2} assumes insensitivity only for a subset of angles ($[-\varphi,\varphi]$ and $[\pi - \varphi, \pi + \varphi]$), while for angles close to orthogonal, the $\cos \theta$ term in Eq. \eqref{Eq:vdpre} guarantees to be in the linear-response regime (right). Green area denotes applicability of the linear-response theory, blue area denotes the insensitive region up the gradient where cells only run, and red area denotes the region where cells immediately tumble. (b) Heat maps of the discrepancy between the predicted limits of the drift velocity in Eq. \eqref{eq:driftSat2} and the observed limits in the simulations. The rotational diffusion used in the simulations is marked with a vertical dashed line. For relatively high receptor activity, Eq. \eqref{eq:driftSat2} does not capture the limits of the drift in the simulations (left). However, for low receptor activity due to inhibition of ligand binding Eq. \eqref{eq:driftSat2} does capture the  limit from simulations for $\varphi \approx \pi/4$ (middle). Note the minimum difference between the theoretical and simulated drift is in the white area of the heat maps as explained by the color bar (right).}
\label{FigLimit}
\end{figure}

\begin{figure}[h!]
\includegraphics[width=\textwidth]{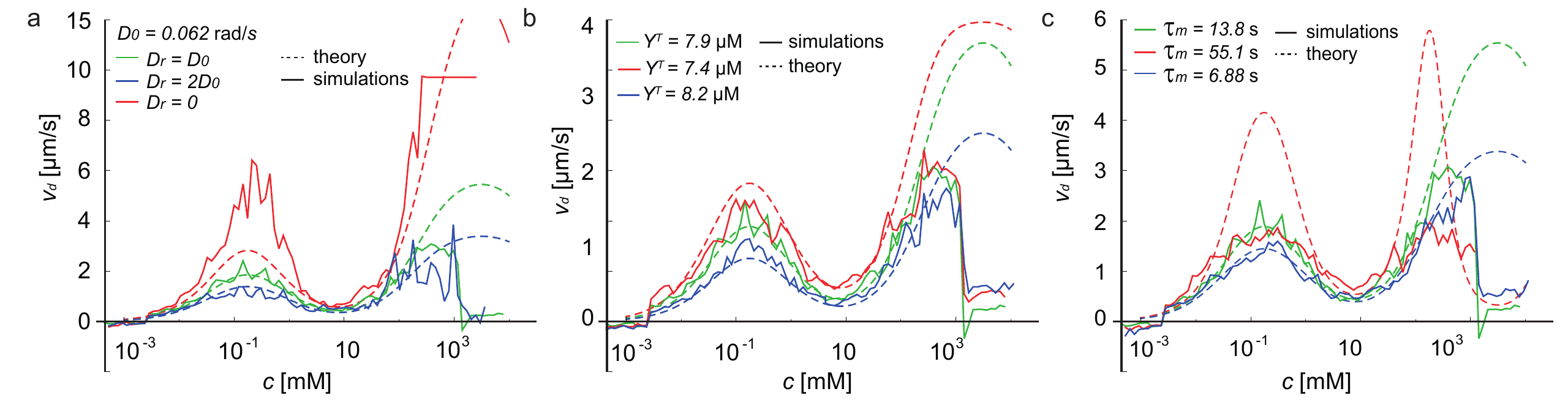}
\caption{{\bf Drift velocity as a function of the attractant concentration for different rotational diffusion coefficients and phenotypes.} Comparison between theory (dashed lines) and simulations (solid lines) for cells swimming in an exponential gradient with $\lambda=0.75 \ \text{mm}^{-1}$. (a) A single phenotype swimming in different media with rotational diffusion coefficient $D_r=D_0$ (green lines), $D_r=2 D_0$ (blue lines), and $D_r=0$ (red lines). $D_0=0.062$ rad$^2$/s. (b) Three different phenotypes that differ for the total level of CheY, $Y^T=7.6 \ \mu$M (green lines), $Y^T=8.0  \ \mu$M (blue lines), and $Y^T=6.0  \ \mu$M (red lines), resulting in a run bias of about 0.85, 0.75 and 0.95, respectively. All the other parameters are the same, rotational diffusion given by $D_0$. (c) Three different phenotypes that differ in adaptation time, $\tau_m=15.5$ s (green lines), $\tau_m=2.15$ s (blue lines), and $\tau_m=214$ s (red lines). All the other parameters are the same with $Y^T=7.6 \ \mu$M and rotational diffusion $D_0$. }
\label{Fig.S1}
\end{figure}

\noindent \textbf{Limits of the drift velocity in simulations} 
\noindent Heuristic explanation: In the following, we estimate the upper bound of the drift velocity in steep gradients at the peaks of the drift (dashed horizontal line in Fig. 2b in the main text), which is about 3.3 $\mu$m/s with rotational diffusion $D_r=0.062$ rad$^2$/s (see main text) and up to 11 $\mu$m/s (close to the run velocity) without rotational diffusion (Fig. \ref{Fig.S1}a). In the case with rotational diffusion, if a cell picks a random direction after a tumble, half of the cells will go down the gradient and immediately tumble again, not contributing to the drift. The other half of the cells will move up the gradient and hence contribute to the drift velocity, leading to $v_d \approx v_0 / (2 \pi) \int_{-\pi/2}^{+\pi/2} \cos \theta \text{d}\theta=v_0/\pi \approx 3.76 \ \mu$m/s, where $v_0=11.81 \ \mu$m/s is the estimated run velocity from experiments (see Fig. \ref{Fig:PreAna}). Note for increased adaptation, the value of the drift velocity will increase since cells going in the $\pm \pi/2$ directions (perpendicular to the gradient) will more likely tumble and thus find a better direction up the gradient, which will increase the drift. Without rotational diffusion, we expect a further increase in the drift velocity. Indeed, all the cells will eventually pick a direction up the gradient and then run straight leading to $v_d \approx 2 v_0 / \pi \approx 7.52 \ \mu$m/s. Hence, arbitrary steep gradients can never increase drift to run velocity $v_0$, i.e. the ballistic limit can never be reached.  

Quantitative explanation: Our analytical theory, as stated before, fails when the change of concentration triggers a change of receptor activity outside the linear-response regime. Without assuming linear response, we can imagine that when a cell moves up a very steep gradient the rate of tumbling is set by rotational diffusion only. Assuming $D_r \gg M\exp\left[ - \left(20 - 40 \frac{Y^p}{Y^p+3.06}\right)\right]/t_r$ in Eq. S7, we obtain $k_{r\rightarrow t}=D_r$ in two dimensions. In contrast, when a cell moves down the gradient we assume that it instantaneously tumbles. Since cells running up the gradient keep running without tumbling, cells going down the gradient contribute to the overall tumble time (Fig. \ref{FigLimit}a middle). With these assumptions, the limiting drift velocity from Eq. \eqref{Eq:vdpre} becomes
\begin{align}
\label{eq:driftSat1}
v^{D_r}_d =  \frac{ \int_{-\pi/2}^{\pi/2} \frac{1}{D_r} \cos(\theta) \ \text{d} \theta } { \int_{-\pi/2}^{\pi/2} \frac{1}{D_r} \ \text{d} \theta  +  \int_{\pi/2}^{3\pi/2} \frac{1}{\left. k_{t \rightarrow r}\right|_0} \ \text{d} \theta } v_0  = \frac{2 \left. k_{t \rightarrow r}\right|_0}{ \pi D_r + \pi \left. k_{t \rightarrow r}\right|_0} v_0 .
\end{align}
Hence, without rotational diffusion we have $v^{D_r=0}_d = \frac{2}{\pi}v_0 \approx 7.5 \ \mu$m/s (which approximately matches the Tar peak in Fig. \ref{Fig.S1}a). 
However, with the rotational diffusion coefficient $D_r=0.062$ rad$^2$/s used in simulations, the theoretical limiting velocity in Eq. \eqref{eq:driftSat1} is $v^{0.062}_d = 7.36  \ \mu$m/s which is not able to capture the drift limit of $\approx 3.3 \ \mu$m/s shown in Fig. 2b of the main text (peak drift in the simulations).
This might be the case because cells do not have saturated responses over all swim angles. Take for instance a trajectory going perpendicular to a very steep gradient, where $\cos \theta$ and hence the experienced gradient are small. For such trajectories, the cell's drift is well captured by the linear-response regime in which cells can run and tumble. Thus, the assumption that cells only run up the gradient and only tumble down the gradient can still be true but only for a smaller set of angles in which the experienced gradient saturates the response (Fig. \ref{FigLimit}a right). Specifically, we assume that for angles $\in [-\varphi, \varphi]$ cells run only and for angles $\in [\pi - \varphi , \pi+\varphi]$ cells tumble only, while for the remaining angles cells run \textit{and} tumble. Hence, Eq. \eqref{eq:driftSat1} can further be generalized for this case and becomes
\begin{align}
\label{eq:driftSat2}
v^{D_r}_d  =  \frac{ \int_{0}^{\varphi} \frac{1}{D_r} \cos(\theta) \ \text{d} \theta + \int_{\varphi}^{\pi-\varphi} \tau_r(\theta) \cos(\theta) \ \text{d} \theta } { \int_{0}^{\varphi} \frac{1}{D_r} \ \text{d} \theta  + \int_{\varphi}^{\pi-\varphi} \tau_r(\theta) \ \text{d} \theta +  \int_{\varphi}^{\pi} \frac{1}{\left. k_{t \rightarrow r}\right|_0} \ \text{d} \theta } v_0 =  
\frac{ \left[ \sin (\varphi)   + D_r \int_{\varphi}^{\pi-\varphi} \tau_r(\theta) \cos(\theta) \ \text{d} \theta \right] \left. k_{t \rightarrow r}\right|_0} { \varphi \left. k_{t \rightarrow r}\right|_0 + D_r \left. k_{t \rightarrow r}\right|_0 \int_{\varphi}^{\pi-\varphi} \tau_r(\theta) \ \text{d} \theta +  (\pi-\varphi) D_r  }  v_0 , \end{align}
where $\tau_r(\theta)$ is given by  Eq. \eqref{Eq:taur}. Although the limiting drift velocity in Eq. \eqref{eq:driftSat2} is still based on strong assumptions it is able to reduce the gap between the predicted limit of the drift velocity and the observed limit in the simulations. Indeed, for a low average receptor activity and a slightly higher diffusion coefficient our equation approaches $3.3 \ \mu$m/s (Fig. \ref{FigLimit}b).

\noindent \textbf{Additional comparison between theory and simulations.} 
Fig. \ref{Fig.S1} shows the analytical drift velocity and the drift from simulations for different rotational diffusion coefficients and phenotypes. 
As expected for large drift there are discrepancies between theory and simulations in all three panels. In panel a, the drift from simulations is above the drift from the analytical theory, while in panels b and c simulations are below the theory result (especially in panel c). We would expect that our theory results are above our simulations results for large drift as our analytical theory does not capture nonlinear effects and memory (our horizontal limit in Fig. 2b of the main text). The reason for the drift from simulations being above the drift from our theory (in panel a) might have to do with the issue that the analytical theory overestimates tumbles by assuming that the linear-response regime is valid. In contrast, in our simulations all cells eventually align themselves with the gradient in absence of rotational diffusion and then stop tumbling due to receptor saturation (inset of Fig. 2a of the main text).

\section{Design of the experiments}

\begin{figure}[h!]
\centering
\includegraphics[width=\textwidth]{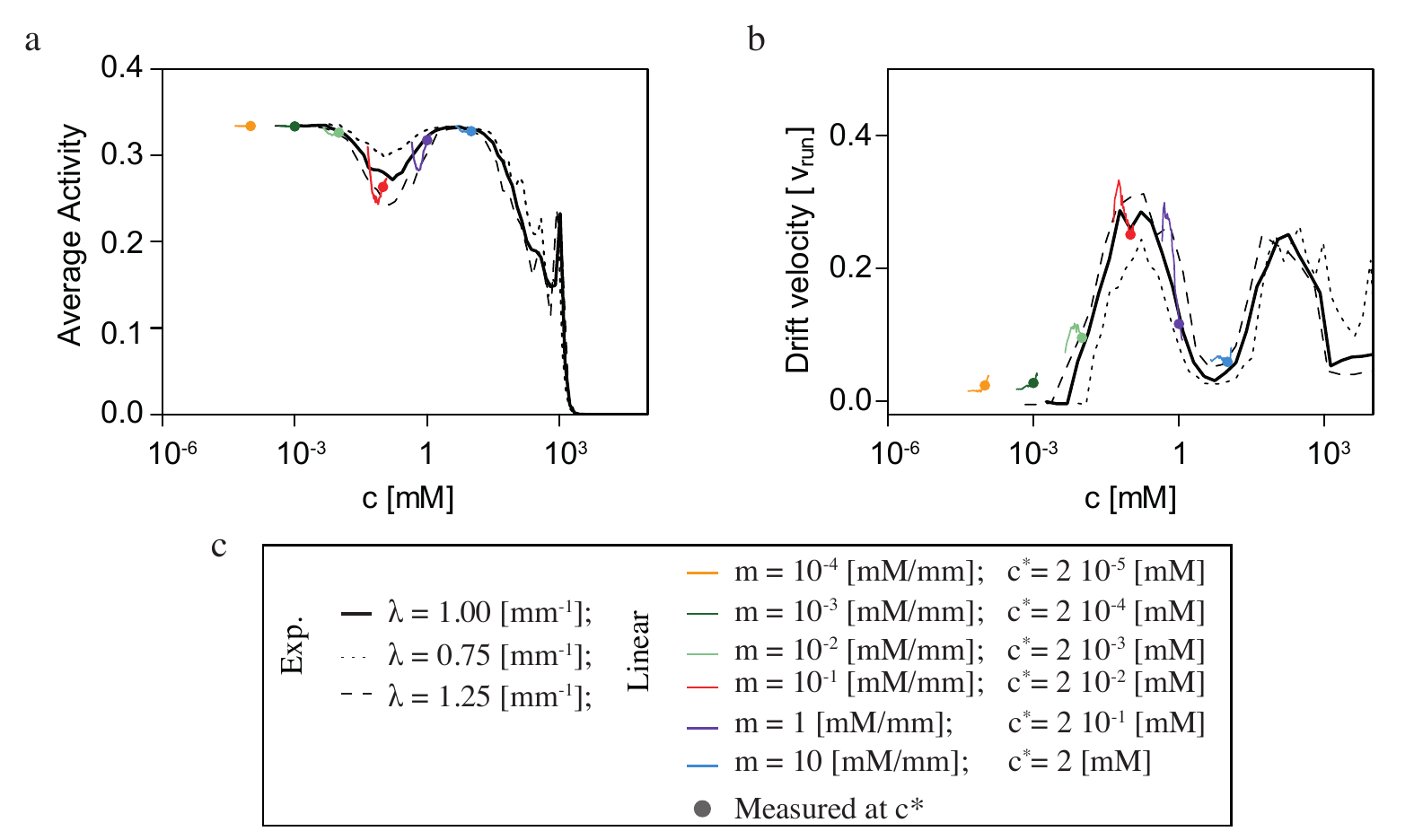}
 \caption{{\bf Swimming behavior in exponential and linear gradients.} 
(a) The average activity as a function of the concentration from simulations of 100 cells. 
(b) The average drift velocity as a function of the concentration from simulations of 100 cells. Both the average activity and the drift velocity in linear gradients match the respective average activity and drift in exponential gradients at $c^*(y^*)$. 
(c) Legend of panels (a) and (b), where $\lambda$ is the exponent in the exponential gradient and $m$ is the slope of the linear gradient made to match the condition $m=\lambda c$ at $c=c^*$.
\label{Fig:expDes}}
\end{figure}

\noindent Equations \eqref{eq:driftN} and \eqref{Eq:Aav} together predict that the average drift of a population of bacteria swimming in a (shallow) gradient has two peaks of high drift (Figs. 2 and 3 in main text). 
However, firstly for concentrations higher than 10 mM, other physical parameters such as the osmolarity of the buffer solution and its viscosity are sufficiently strongly varying along the gradient to affect the chemotactic behaviour in the experiments. For this reason, the experiments we proposed focus on the first peak only.   
Secondly, it is important to question whether experiments in linear gradients provide insights about the swimming behavior in exponential gradients. 
In Eq. \eqref{eq:driftN}, the term $\partial F/\partial c \nabla c$ is the product of methylation-independent receptor activity and the local relative gradient, and this term depends on the concentration and gradient only. 
While $\partial F/\partial c \nabla c$ can be set to be the same in exponential and in linear gradients, $K(A)$ depends on the history and hence might be different in different gradients. 
For instance, by focusing on concentration $c^*$ a linear gradient $c = m y + c_0$ and an exponential gradient $c=Z e^{\lambda (y-y_0)}$ have the same relative gradient if $m=\lambda c^*$. To determine whether the drift is the same in linear and exponential gradients under these conditions, despite potential effects from memory, we perform simulations. 
As shown in Fig. \ref{Fig:expDes} both the average drift and the average activity in linear gradient at different $c^*$ match the average drift and average activity in exponential gradient, giving us the confidence to carry out the experiments. 

\section{Experimental data acquisition and analysis}

\noindent \textbf{Cell growth.} \textit{Escherichia coli} strain MG1655 was grown overnight in Trypton Broth (TB) at 30$^o$C, diluted 1/100 in 10 ml fresh TB in a flask, and grown for 3.5 h at 34$^o$C under vigorous shaking until an optical density at 600 nm of OD600 = 0.7. The cells were then washed 3 times in motility buffer (MB - (10 mM KPO$_4$, pH 7, 0.1 mM EDTA, 67 mM NaCl), kept at 4$^o$C for 20 minutes to reduce metabolic activity and diluted 30 times in MB. The cell suspension was then supplemented with $0.25\%$ \textit{wt} Glucose as an energy source.  Aliquots of 80 $\mu$l of the resulting cell suspension were mixed with either 20 $\mu$l of plain MB or 10 $\mu$l of a 100$\mu$M suspension of fluorescein in MB plus 10 $\mu$l of various dilutions of $\alpha$-D,L-methylaspartic acid (MeAsp) in MB (the pH of which had been adjusted to 7), in order to create cell suspensions supplemented with MeAsp at concentrations 20mM, 2mM, 200 $\mu$M, 20 $\mu$M, 2$\mu$M and 0 $\mu$M.

\ \\

\noindent \textbf{Chemotactic measurement.}
The chemotaxis experimental device was produced using standard photolithography. It consists of 2 large reservoirs (2 cm$^2$ x 50 $\mu$m height) linked by a rectangular channel (2 mm long x 1 mm wide x 50 $\mu$m height). One of the reservoirs is filled with a MeAsp containing cell suspension while the other contains the $c_{\text{MeAsp}} = 0\mu$M cell suspension. A linear gradient of MeAsp forms in the channel, which becomes quasi-static in about 1 hour, to which the cells respond chemotactically. The sample is observed at 10x magnification (NA=0.3) under phase contrast, or green fluorescence (excitation 470/40 nm, emission 525/50 nm, for gradient measurement), wide field illumination. The motion of the cells was recorded using a Mikrotron 4CXP CMOS camera (1 px = 0.7 $\mu$m, 1024 x 1024 px$^2$ field of view, 30 frames per second, 100 s long movie) in the middle of the channel, 2, 3 and 4 hours after the chamber had been filled. The gradient of fluorescein was then measured, as a proxy for the gradient of MeAsp (since fluorescein and MeAsp have very similar diffusion coefficients \cite{Ahmed10}, using an ANDOR Zyla sCMOS camera (1 px = 0.65 $\mu$m, 2048 x 2048 px$^2$ field of view) by taking 10 fluorescence pictures of the center of the channel (500 ms exposure), as well as 10 pictures in the MeAsp + fluorescein reservoir, at least 500 $\mu$m away from the channel entrance, for normalization purposes. The alignment of the two cameras was measured using featured objects, in order to match assign a concentration to each pixel in the tracking movie.

\ \\

\begin{figure}[h!]
\includegraphics[width=\textwidth]{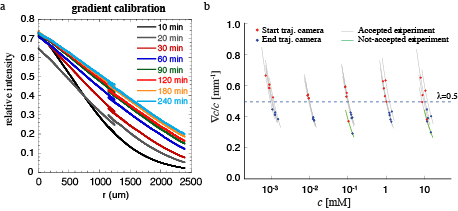}
\caption{{\bf Gradient calibration and measurements.}
(a) Gradient changes in time, but is considered stable after 2 hours. Note that the gradient was measured in 2 subsequent steps, due to the size of the camera chip, explaining the 2 half curves. (b) Measured relative gradients as a function of the concentration in the chamber. Note that the gradient camera is longer than the tracking camera, thus the start and end of the tracking camera are highlighted with red and blue dots, respectively. Two experiments have been disregarded since the trajectory camera field had a relative gradient different than the target gradient ($\lambda=0.5$ $\text{mm}^{-1}$).}
\label{Fig:gradTune}
\end{figure}

\noindent \textbf{Data analysis: gradient measurement.} 
The averages of the 10 pictures both in the channel and in the reservoir were computed. The average channel picture was divided pixel by pixel by the reservoir picture to flatten out the heterogeneities in illumination. The gradient profile was then computed along the length of the channel by averaging pixel intensities along its width. Gradient measurements were performed after chemotactic measurements, and the stability of the gradient was checked separately (Fig. \ref{Fig:gradTune}a). All gradient measurements are shown in Fig. \ref{Fig:gradTune}b. 

\ \\

\noindent \textbf{Data analysis: particle tracking.}
The movie of swimming cells in phase contrast illumination was treated as follows. Each image was divided by the temporal average of the field of view to remove background features. Cells were detected using a custom written particle tracking. In short, cells were identified as groups of 8-connected pixels below a user-defined, film specific, threshold. Their position was computed as the intensity-weighted average position of the group of pixel. The cell positions in all frames were linked into trajectories using the standard proximity criterion of \cite{Crocker1996}, with only linking in strictly subsequent frames permitted.  The thus obtained trajectories were then used for further analysis. 

\begin{figure}[h!]
\includegraphics[width=\textwidth]{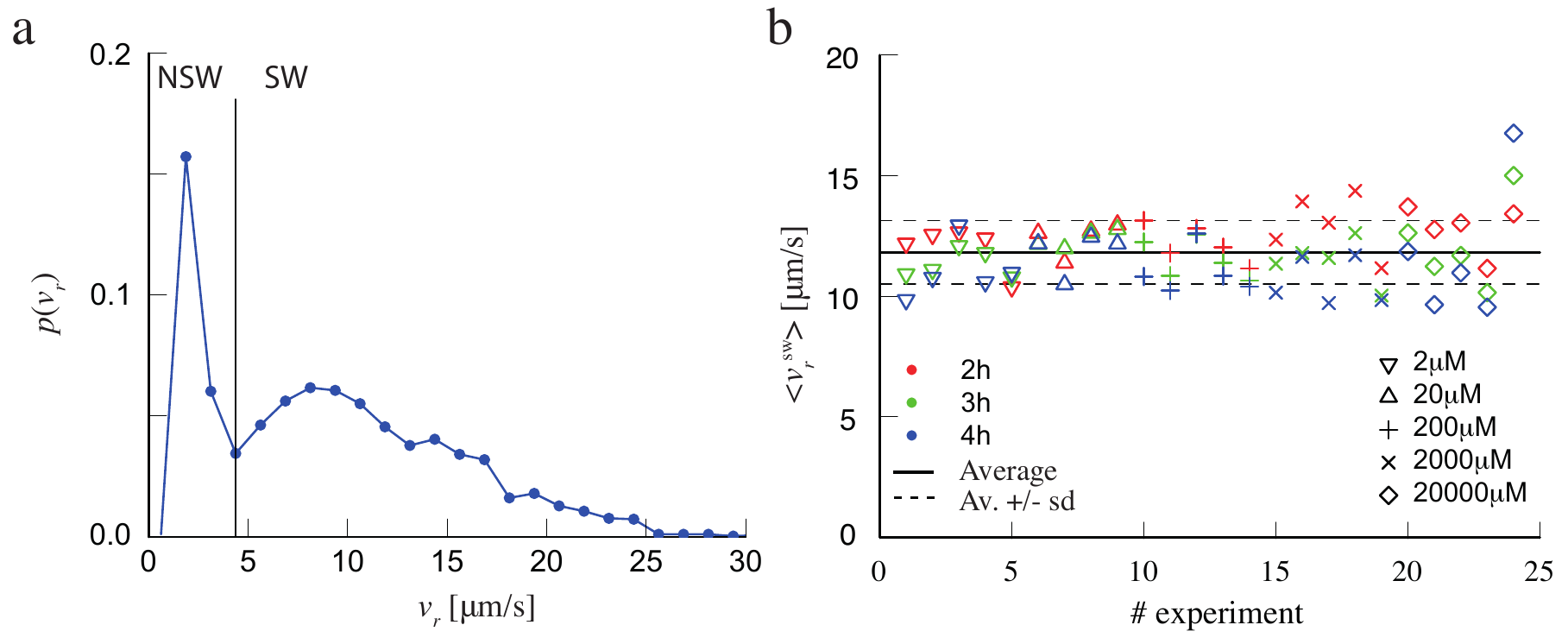}
\caption{{\bf Run velocity and discrimination between swimmers (SW) and non-swimmers (NSW).}
(a) Typical distribution of the recorded particles velocity for one experiment. The distribution of all the particles (blue line and dots) is bimodal, allowing the distinction between non-swimming particles (left of the black solid vertical line) and swimming particles (right of the black solid vertical line). (b) Average run velocity for swimmer only. Each dot correspond to different experiments.}
\label{Fig:PreAna}
\end{figure}

\noindent \textbf{Data analysis: the average speed velocity enables the distinction between swimming cells and non-swimming particles.}
To calculate the drift velocity, it is important to discriminate between swimming bacteria and non-swimming particles, such as bacteria that have lost their flagella or Brownian particles. Only trajectories longer than 1 s where considered, which ensured the observation of at least one run (note that the tumble average time is 0.1 s). 
Furthermore, the velocity distribution of all particles is bimodal, which allows for the discrimination between swimmers and non-swimmers (Fig. \ref{Fig:PreAna}a). The average run velocity for the swimming bacteria is $v_0=11.8 \pm 1.3 \ \mu$m/s (Fig. \ref{Fig:PreAna}b). 

\ \\

\begin{figure}[h!]
\includegraphics[width=\textwidth]{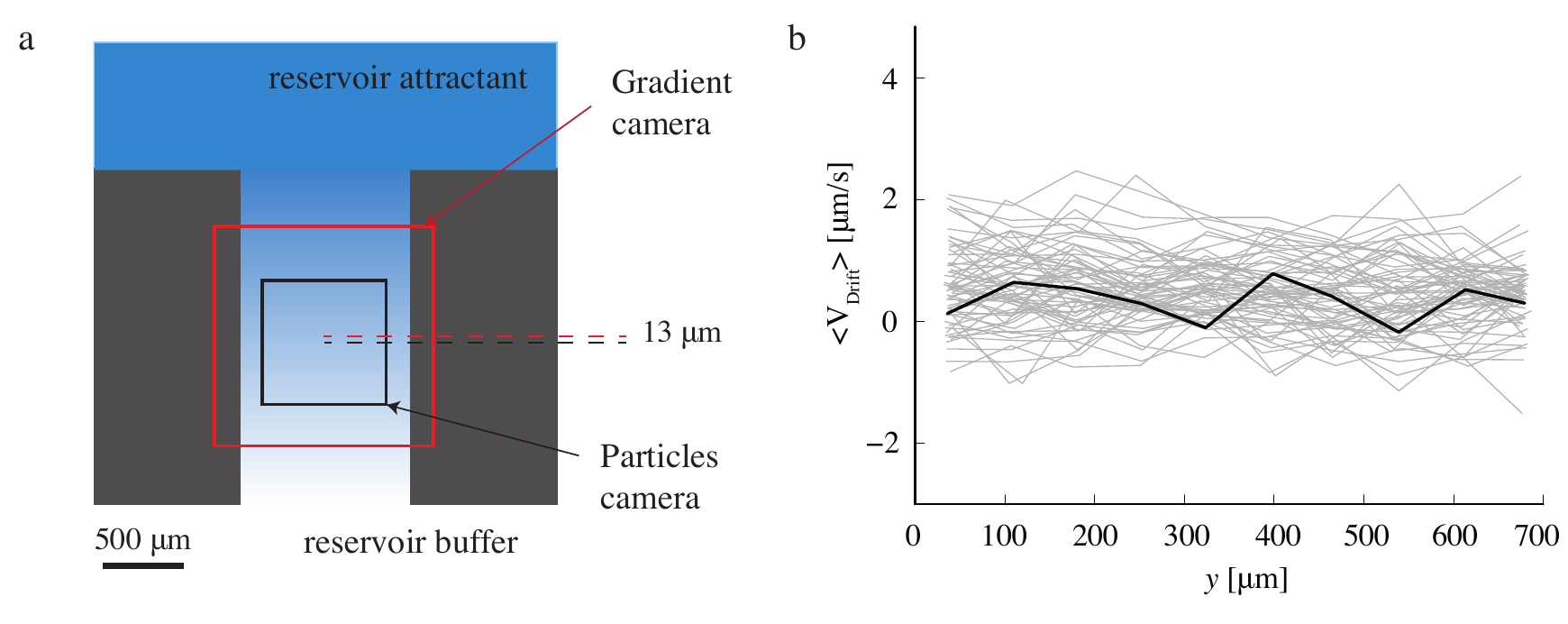}
\caption{{\bf Experimental setup and drift velocity in the camera field.}
(a) Schematic of the experimental setup. (b) Average drift velocity as a function of the location in the particles camera for all the experiments (grey lines). One experiment is shown by a black line to highlight the constant noisy trend. }
\label{fig:vd_y}
\end{figure}

\noindent \textbf{Data analysis: Evaluation of the drift velocity.} The drift velocity was calculated using the tracking trajectories. 
Since the qualification of the drift velocity as a function of the position along the gradient axes did not show a statistical relevant decreasing velocity (see Fig. \ref{fig:vd_y}), the drift velocity is calculated by averaging over the whole camera view field. In order to account for an eventual flow in the chamber, the drift velocity of non-swimmers (supposedly zero, in absence of flow) has been subtracted from the drift velocity of the swimmers, $v^{\text{Exp}}_d=v^{\text{SW}}_d-v_d^{\text{NSW}}$. 

\ \\

\section{Information gain, accuracy of sensing and thermodynamic costs of chemotaxis}

Our theory and experiments were able to explain the fine structure of cell behavior in trajectories based on the well-known chemotactic pathway in \textit{E. coli}. However, at a more fundamental level, how do regions of high drift emerge from information gain and energy consumption, just considering trajectories (and ignoring most \textit{E. coli}-specific details)? Previous application of information theory to \textit{E. coli} chemotaxis indicated that the assumption of maximal information transmission (mutual information) leads to maximal drift \cite{Micali14}. This suggested a potential new design principle in the bacterial sensory systems, i.e. that maximizing information transmission optimizes cell behavior. 
However, this theory had a number of limitations, including the restriction to instantaneous responses, thus neglecting any history dependency and memory effects \cite{Micali16}. Calculating the mutual information based on trajectories is difficult due to the high-dimensional space of trajectories, requiring additional approximations \cite{TosttenWolde09}. 
In the following, we illustrate in a heuristic way the connection between drift, information gain, sensing accuracy, and energy dissipation.

\subsection{Information gain} 

To quantify information gain, we used the Kullback-Leibler divergence (KL). 
In general, the KL-divergence between two probabilities of observing an event $s$ is defined by $\text{KL}\left(p_s,q_s \right) := \sum_s p_s \log_2 p_s/q_s$. Note that the KL-divergence is defined only if for all the events $s^*$ with probability $q$ equal zero, ($q_{s^*}=0$) probability $p$ is also equal to zero ($p_{s^*}=0$). 

Here, in order to evaluate the information gain (KL-divergence) along a trajectory at time $t$ between the probability of runs and tumbles from simulations in exponential gradient, $p_{s,t}$, relative to the probability from adapted cells without gradient, $q_s$ 
\begin{align} 
\text{KL}(p_{s,t}, q_s) = \sum_{s=\{ \text{run, tumble} \} }  p_{s,t} \log \left( \frac{p_{s,t}}{q_s} \right),
\end{align}
using our simulations. $\text{KL}(p_{s,t}, q_s)$ estimates the information lost once the probability of runs and tumbles in absence of chemotaxis is used to explain the behavior of the cells.  

Note that $p_{s,t}$ is the probability of running (or tumbling) at the time $t$ along the trajectory, while $q_s$ is the average probability of running (or tumbling) extracted from simulations without a gradient. KL is calculated for each trajectory. At a given concentration $c$ experienced by the cells in the box, KL is then averaged over many trajectories. Indeed, $\text{KL}(p_{s,t}, q_s)$ peaks at the concentration where the drift is maximal (Fig. \ref{fig3}a,b, blue line). Does information gain also lead to more accurate sensing of the gradient?

\subsection{Lower bound on the accuracy of sensing} 

\begin{figure}[th!]
\includegraphics[width=0.6\textwidth]{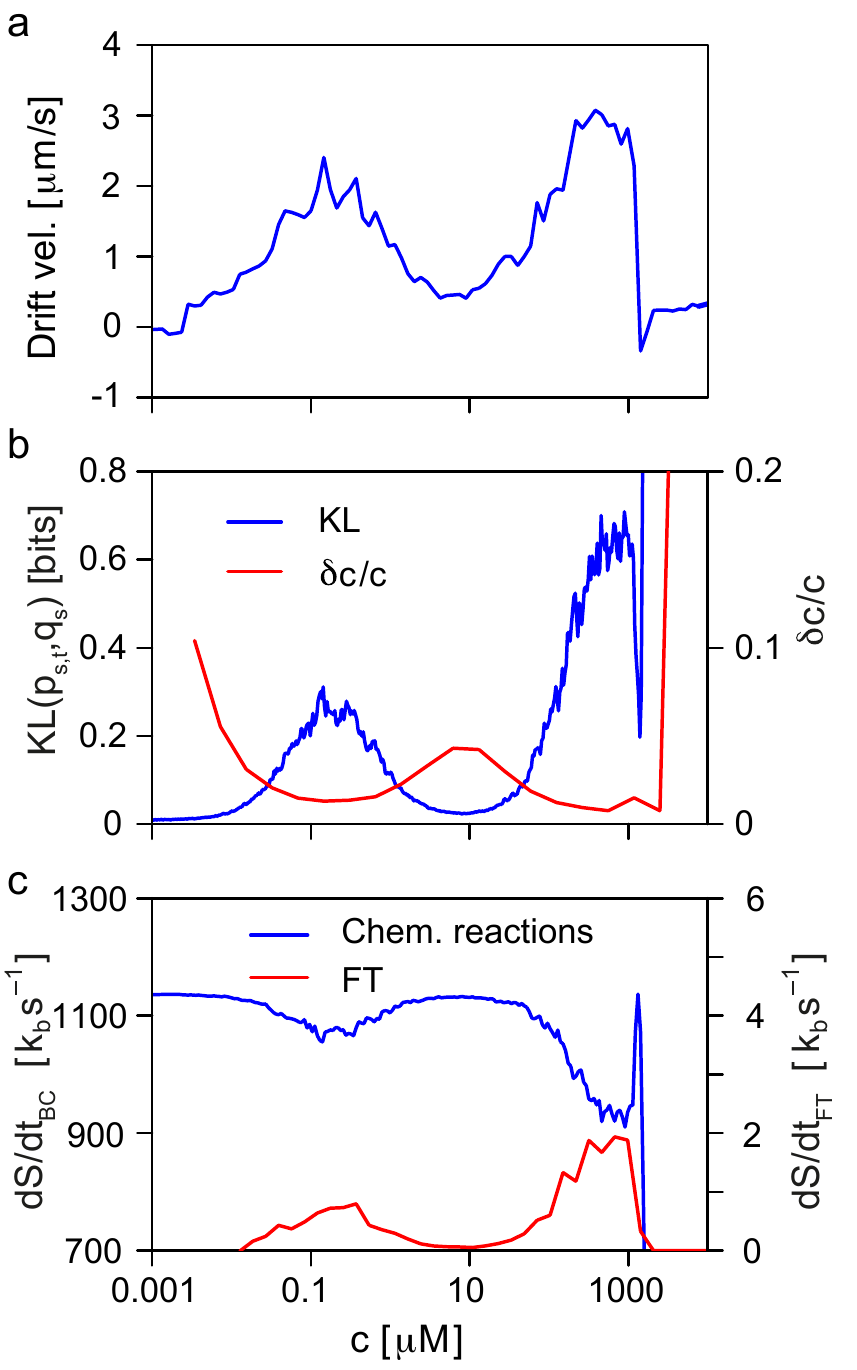}
\caption{{\bf Connection between behavior, information gain, and energy dissipation.}
(a) Average drift velocity.
(b) Kullback-Leibler divergence as a measure of information gain (blue) and uncertainty of sensing from Cram\'er-Rao bound (red). 
(c) Actual entropy production from biochemistry (blue) and lower bound from fluctuation theorem (red). 
Note the different scales.
}
\label{fig3}
\end{figure}

To quantify the sensing accuracy, we propose an approach inspired by the Fisher information from estimation theory \cite{BrunelNadal}. The Fisher information focused on the likelihood, given by the conditional probability of observing an outcome given a certain parameter. Indeed, this definition can be interpreted as the curvature of the log-likelihood function and as a result the Fisher information cannot be estimated from a single event. Furthermore, it estimates a lower bound of the uncertainty of sensing, via the Cram\'er-Rao bound 
\begin{equation} 
\frac{\delta c} {c} \ge \frac{1}{\sqrt{c \mathcal{F}(c) }}. 
\label{Eq:CR}
\end{equation}
In case of swimming bacteria performing chemotaxis, we are interested in a quantity similar to Fisher information that can be defined along a swimming trajectory. In this case, the likelihood is given by the conditional probability of the cell to run or tumble given the external concentration and the cell's history, i.e. its trajectory. Thus for a trajectory, the stringent definition of Fisher information is lost. However, we can still computationally evaluate a Fisher-like quantity which estimates the information of the concentration sensed by a cell swimming up a chemical gradient over individual trajectories. 
In particular, this Fisher-like information along a trajectory at time $t$ is defined by 
\begin{align}
F_{t}(c) :=  \sum_{s=\textit{\{run,tumble\}}} p_{s,t} \left( \frac{ \log(p_{s,t+\delta t})- \log(p_{s,t}) }{c_{t+\delta t} - c_{t}} \right)^2 .
\end{align} 
Further extending the parallelism between Fisher information, $\mathcal{F}$, and Fisher-like information, $F$, we evaluate the equivalent of the Cram\'er-Rao bound for $F$. 
Interestingly, when averaged over many trajectories at a fixed concentration, the lower bound in Eq. \eqref{Eq:CR} dips at concentrations at which the drift and the KL divergence peak (Fig. \ref{fig3}b, red line), showing that, in principle, cells bias their movement in regions where they have higher accuracy of sensing. Note that to reach the Cram\'er-Rao bound even for the Fisher information, an efficient estimator is required. Here, the Fisher-like information is calculated over two data points, and thus the lower bound may not be reached.

\subsection{Energetic costs of chemotaxis} 

Here, we estimate the energetic cost by the entropy production from the biochemical reactions.  
Given a biochemical reaction $\ce{ A <=>[k^+][k^-] B }$ with the forward rate $r^+=k^+ [A]$ and the backward rate $r^-=k^- [B]$, we can write the following differential equation $\text{d}A/\text{d}t = k^- [B] - k^+ [A]$, and the entropy production for a biochemical (BC) reaction is \cite{Qian05, Qian07} 
\begin{equation}
\frac{ \text{d} S_{\text{BC}} }{\text{d}t} = k_\text{B} (r^+-r^-) \log \left( \frac{r^+}{r^-} \right), 
\label{Eq:dSbc}
\end{equation}
where $k_\text{B}$ is the Boltzmann constant. Here, assuming similar viscous drags and hence rotational motor speeds for both runs and tumbles \cite{Berg2003motors}, the proton flux through the motor and hence the energetic costs are assumed to remain constant and independent of the swimming behavior.
The biochemical reactions considered are the methylation of receptors and phosphorylation of the response regulator CheY. 

\ \\

\noindent The phosphorylation of CheY is governed by the reaction
\begin{align*}
&\ce{ Y + \text{ATP} <=>[Ak_y^+][Ak_y^-]} Y_p + \ce{\text{ADP} <=>[Zk_z^+][Zk_z^-] Y + \text{ADP} + \text{P}_i }, 
\end{align*}
where $A$ is the receptor activity, $Z$ is the CheZ concentration, $[\text{ATP}]$, $[\text{ADP}]$, $[\text{P}_i]$ are the concentrations of ATP, ADP and inorganic phosphate, respectively; $Y_p$ and $Y=Y^T-Y_p$ are the concentrations of $\text{CheY}_\text{p}$ and CheY, respectively.
The dynamical equation is
\begin{align*}
\frac{\text{d}Y_p}{\text{d}t} = \tilde{k}^+_y A \left( Y^T - Y_p \right) - \tilde{k}^-_y A Y_p - \tilde{k}^+_z Y_p + \tilde{k}^-_z Y, 
\end{align*}
where effective parameters have been used  \cite{VlaLovSou08}. In particular, $\tilde{k}^+_y = k^+_y [\text{ATP}] = 100 \ \mu\text{M}^{-1} \text{s}^{-1}$, $\tilde{k}^-_y = k^-_y [\text{ADP}] = \nu_1 \tilde{k}^+_y$ with $\nu_1 \ll 1$,  $\tilde{k}^+_z = Z k^+_z = 30 \ \text{s}^{-1}$, and  $\tilde{k}^-_z = Z k^-_z [\text{P}_i]= \nu_2 \tilde{k}^+_z$, with $\nu_2 \ll 1$.
Using Eq. \eqref{Eq:dSbc}, the entropy production of the phosphorylation dynamics is
\begin{align}
\nonumber
\frac{\text{d}S^{\text{Y}}_\text{BC}}{\text{d}t} = & k_\text{B} \Big[ \tilde{k}^+_y (Y^T -Y_p) - \tilde{k}^-_y Y_p \Big] A \ \log \left[ \frac{\tilde{k}^+_y (Y^T -Y_p)}{\tilde{k}^-_y  Y_p} \right]  \\
&+  k_\text{B} \Big[ \tilde{k}^+_z Y_p - \tilde{k}^-_z (Y^T -Y_p) \Big] \ \log \left[ \frac{\tilde{k}^+_z Y_p}{\tilde{k}^-_z (Y^T -Y_p)} \right] .
\end{align}  

\ \\

Similarly, the methylation dynamics are governed by the reactions
\begin{align*}
&  \left[ m \right]_0 + \text{SAM} \ce{ <=>[k_R^+][k_R^-] } \left[ m +1 \right]_0 + \text{SAH}   \\
&  \left[ m +1 \right]_1 +  \text{H}_2\text{O} \ce{ <=>[k_B^+][k_B^-] } \left[ m \right]_1 + \text{CH}_3\text{OH}  , 
\end{align*}
where only the inactive sites $[m]_0$ are methylated and only the active site $[m]_1$ are demethylated.
The effective dynamical equation is given by \cite{ClauOleEnd10}
\begin{align*}
\frac{\text{d}m}{\text{d}t} = \left( \tilde{k}^+_R  - \tilde{k}^-_R \right) (1-A) - \left( \tilde{k}^+_B  - \tilde{k}^-_B \right) A^3 , 
\end{align*}
with effective parameter $\tilde{k}^+_R=k^+_R [\text{SAM}]= 0.0069 \ \text{s}^{-1}$, $\tilde{k}^-_R = k^-_R [\text{SAH}] = \nu_3 \tilde{k}^+_R$, $\tilde{k}^+_B = k^+_B [\text{H}_2\text{O}]= 0.12 \ \text{s}^{-1}$, $\tilde{k}^-_B = k^-_B [\text{CH}_3\text{OH}]= \nu_4 \tilde{k}^+_B$, and $[m]_T=10^4$ the total number of methylation sites \cite{ClauOleEnd10, LiHaz04}. 
Hence, using Eq. \eqref{Eq:dSbc}, the entropy produced by the methylation dynamics is
\begin{align}
\nonumber
\frac{\text{d}S^{\text{m}}_\text{BC}}{\text{d}t} = k_\text{B} & \Big( \tilde{k}^+_R - \tilde{k}^-_R \Big) (1-A) [m]_T \ \log{ \left( \frac{\tilde{k}^+_R}{\tilde{k}^-_R} \right) } + \\ 
  & +  k_\text{B} \Big( \tilde{k}^+_B - \tilde{k}^-_B \Big) A^3 [m]_T \ \log{ \left( \frac{\tilde{k}^+_B }{ \tilde{k}^-_B} \right) }.
\end{align}  
For each trajectory at time $t$, we get the information about the activity, $A$, and the phosphorylated CheY, $Y_p$, and thus we can calculate $\left. \frac{\text{d}S}{\text{d}t} \right|_t = \left. \frac{\text{d}S^{\text{Y}}_\text{BC}}{\text{d}t} \right|_t + \left. \frac{\text{d}S^{\text{m}}_\text{BC}}{\text{d}t} \right|_t$. The average over multiple trajectories at a given concentration $c$ is plotted in Fig. \ref{fig3}c, (blue line).


\subsection{Lower limit on entropy production from fluctuation theorem} 

The Evans-Searles fluctuation theorem (FT) focuses on the probability $p(\Omega_t)$ of observing a functional $\Omega_t=\Omega[x(t)]$ along a trajectory of length $t$. Such a functional is normally the dimensionless dissipated energy. The theorem states 
\begin{equation}
\frac{ p\left(\Omega_t=\sigma\right) }{ p\left(\Omega_t=-\sigma\right) } = e^{\sigma},
\label{Eq:FT}
\end{equation}
which expresses the ratio of the probabilities of observing trajectories of duration $t$ with values of the dissipation function $\Omega_t$ being $\sigma$ and $-\sigma$, respectively. 
Although the dissipated energy would depend on the system, under time-reversible mechanics, any trajectory with $\Omega_t=\sigma$ has a time-reversed trajectory (`anti-trajectory') with $\Omega_t=-\sigma$ \cite{Sevick08}. This means that Eq. \eqref{Eq:FT} can be interpreted as a relation between trajectories ($\Gamma$) and anti-trajectories ($\bar{\Gamma}$), $p_\Gamma / p_{\bar{\Gamma}} = \exp(\delta S_\text{FT})$. 
Given $\Gamma$ and $\bar{\Gamma}$, the probabilities of the cell to run in a time $\delta t$ can be calculated from the model of the chemotaxis pathway for both the trajectory and anti-trajectory ($p^r_{\Gamma_t}$ and $p^r_{\bar{\Gamma}_t}$, respectively). Thus, $\frac{ \delta S_{\text{FT}}}{\delta t} = \log \left( p^i_{\Gamma_t} / p^i_{\bar{\Gamma}_t} \right)$, to evaluate the entropy production along trajectories at time $t$, in units of $k_\text{B}$. Here, $p^i$ is the probability of continued running, if the cell is in the run state, or the probability of continued tumbling if the cell is in the tumble state. 

We found that the lower bound on the entropy production, once averaged over multiple trajectories at a given concentration, shows again a peak in the region of high drift (Fig. \ref{fig3}c, red line). 
This also shows that the actual entropy production based on biochemistry is orders of magnitude higher, and it is interesting to wonder if all this energy consumption is needed for efficient chemotaxis.  

\ \\

\noindent \textbf{Calculation of anti-trajectories.} The anti-trajectories are computed by reversing the trajectory and the dynamics. Given a trajectory of a cell $i$, 
\[ \Gamma=\Gamma(i,t, x_t, y_t, \theta_t, c_t, A_t, m_t, \text{CCWb}_t, \text{nCW}_t, \text{RoT}_t), \] 
where $t$ is the time, ($x_t, y_t$) the position, $\theta_t$ the orientation, $c_t$ the sample external concentration of attractant, $A_t$ the average receptor activity, $m_t$ the average methylation level, $\text{CCWb}_t$ the CCW bias, $\text{nCW}_t$ the number of motors rotating CW and RoT$_t$ the state of the cell,
the anti-trajectory is  
\begin{align*} 
\bar{\Gamma}=\bar{\Gamma} \big( i, \ t^r=T-t, \ & x^r_t=x_{T-t},  \  y^r_t=y_{T-t}, \ \theta^r_t=\theta_{T-t}-\pi, \ c^r_t=c_{T-t}, \\
& A^r, m^r, \text{CCWb}^r, \text{nCW}^r_t=\text{nCW}_{T-t}, \text{RoT}_t^r=\text{RoT}_{T-t} \big), 
\end{align*} 
where $T$ is the time at the end of the simulation. 
$A^r$, $m^r$, $\text{CCWb}^r$ are calculated by assuming the initial state to be $\bar{\Gamma}_0=\bar{\Gamma}_0(i,0, x_{T}, y_{T}, \theta_{T}-\pi, c_{T}, A_T, m_T, \text{CCWb}_T, \text{nCW}_{T}, \text{RoT}_{T})$, and inverting the dynamics, i.e. applying Eqs. \eqref{Eq:A}, \eqref{Eq:m}, \eqref{Eq:kccwTOcw} and \eqref{Eq:kcwTOccw} to the reverse quantities $A^r$, $m^r$, $c^r$.
Note, that the probability $p^i_{\Gamma_t}$ of $i=\text{running or tumbling}$ of a trajectory $\Gamma$ at time $t$ is given by 
\begin{align}
p^i_{\Gamma_t} =
\begin{cases}
 \left(1-k^t_{\text{CCW}\rightarrow\text{CW}} \delta t \right)^3 \ \ \ \ \ \ \ \ \ \ \ \  \ \ \ \ \ \ \ \ \ \ \ \  \ \ \ \ \ \ \ \  \ & \text{if nCW=0} \\
 1-\left(1-k^t_{\text{CCW}\rightarrow\text{CW}} \delta t \right)^2 \left(k^t_{\text{CW}\rightarrow\text{CCW}} \delta t\right)  &\text{if nCW=1} \\
 1-\left(1-k^t_{\text{CCW}\rightarrow\text{CW}} \delta t\right) \left(k^t_{\text{CW}\rightarrow\text{CCW}} \delta t\right)^2 &\text{if nCW=2} \\
 1-\left( k^t_{\text{CW}\rightarrow\text{CCW}} \delta t \right)^3  & \text{if nCW=3} \\
\end{cases}
\label{Eq:pRun}
\end{align}
and $p^i_{\bar{\Gamma}_{t^r}}$ is given by Eq. \eqref{Eq:pRun} for the reverse trajectory.



\begin{thebibliography}{52}
\providecommand{\url}[1]{\texttt{#1}}
\providecommand{\urlprefix}{ }

\bibitem[Sourjik and Wingreen(2012)]{SourjikWingreenRev}
Sourjik, V., and N.~S. Wingreen, 2012.
\newblock Responding to chemical gradients: bacterial chemotaxis.
\newblock \emph{Curr. Opin. Cell. Biol.} 24:262--268.

\bibitem[Waite et~al.(2016)Waite, Frankel, Dufour, Johnston, Long, and
  Emonet]{EmonetWaite16MSB}
Waite, A.~J., N.~W. Frankel, Y.~S. Dufour, J.~F. Johnston, J.~Long, and
  T.~Emonet, 2016.
\newblock Non-genetic diversity modulates population performance.
\newblock \emph{Molecular Systems Biology} 12:895.

\bibitem[SIt()]{SItext}
See Supplementary Material for the details on calculations, analysis,
  and software design.

\bibitem[Reneaux and Gopalakrishnan(2010)]{Reneaux_10}
Reneaux, M., and M.~Gopalakrishnan, 2010.
\newblock Theoretical results for chemotactic response and drift of {E.} coli
  in a weak attractant gradient.
\newblock \emph{J Theor Biol} 266:99--106.

\bibitem[Celani and Vergassola(2010)]{CelVerg2010}
Celani, A., and M.~Vergassola, 2010.
\newblock Bacterial strategies for chemotaxis response.
\newblock \emph{Proc. Natl. Acad. Sci. U.S.A.} 107:1391--1396.

\bibitem[Jiang et~al.(2010)Jiang, Ouyang, and Tu]{Tu10}
Jiang, L., Q.~Ouyang, and Y.~Tu, 2010.
\newblock Quantitative modeling of {E}scherichia coli chemotactic motion in
  environments varying in space and time.
\newblock \emph{PLoS Comput. Biol.} 6:e1000735.

\bibitem[Si et~al.(2012)Si, Wu, Ouyang, and Tu]{Tu12}
Si, G., T.~Wu, Q.~Ouyang, and Y.~Tu, 2012.
\newblock Pathway-based mean-field model for Escherichia coli chemotaxis.
\newblock \emph{Phys. Rev. Lett.} 109:048101.

\bibitem[Dufour et~al.(2014)Dufour, Fu, Hernandez-Nunez, and Emonet]{Dufour14}
Dufour, Y.~S., X.~Fu, L.~Hernandez-Nunez, and T.~Emonet, 2014.
\newblock Limits of feedback control in bacterial chemotaxis.
\newblock \emph{PLoS Comput. Biol.} 10:e1003694.

\bibitem[Colin et~al.(2014)Colin, Zhang, and Wilson]{Colin14}
Colin, R., R.~Zhang, and L.~Wilson, 2014.
\newblock Fast, high-throughput measurement of collective behaviour in a
  bacterial population.
\newblock \emph{J. R. Soc. Interface} 11:20140486.

\bibitem[Kalinin et~al.(2009)Kalinin, Jiang, Tu, and Wu]{KalTuWu09}
Kalinin, Y.~V., L.~Jiang, Y.~Tu, and M.~Wu, 2009.
\newblock Logarithmic sensing in \emph{Escherichia coli} bacterial chemotaxis.
\newblock \emph{Biophys. J.} 96:2439--2448.

\bibitem[Shimizu et~al.(2010)Shimizu, Tu, and Berg]{ShimTuBerg10}
Shimizu, T.~S., Y.~Tu, and H.~C. Berg, 2010.
\newblock A modular gradient-sensing network for chemotaxis in
  \emph{Escherichia coli} revealed by responses to time-varying stimuli.
\newblock \emph{Mol Syst Biol} 6:382.

\bibitem[Clausznitzer et~al.(2014)Clausznitzer, Micali, Neumann, Sourjik, and
  Endres]{Micali14}
Clausznitzer, D., G.~Micali, S.~Neumann, V.~Sourjik, and R.~G. Endres, 2014.
\newblock Predicting chemical environments of bacteria from receptor signaling.
\newblock \emph{PLoS Comput. Biol.} 10:e1003870.

\bibitem[Wong-Ng et~al.(2016)Wong-Ng, Melbinger, Celani, and
  Vergassola]{Vergassola_PCB16}
Wong-Ng, J., A.~Melbinger, A.~Celani, and M.~Vergassola, 2016.
\newblock The Role of Adaptation in Bacterial Speed Races.
\newblock \emph{PLOS Comput. Biol.} 12:e1004974.

\bibitem[Long et~al.(2017)Long, Zucker, and Emonet]{EmonetLong17PlosCB}
Long, J., S.~W. Zucker, and T.~Emonet, 2017.
\newblock Feedback between motion and sensation provides nonlinear boost in
  run-and-tumble navigation.
\newblock \emph{PLoS computational biology} 13:e1005429.

\bibitem[Masson et~al.(2012)Masson, Voisinne, {Wong-Ng}, Celani, and
  Vergassola]{MassonCelaniVerg2012}
Masson, J.-B., G.~Voisinne, J.~{Wong-Ng}, A.~Celani, and M.~Vergassola, 2012.
\newblock Noninvasive inference of the molecular chemotactic response using
  bacterial trajectories.
\newblock \emph{Proc. Natl. Acad. Sci. U.S.A.} 109:1802--1807.

\bibitem[Vladimirov et~al.(2008)Vladimirov, L\o{}vdok, Lebiedz, and
  Sourjik]{VlaLovSou08}
Vladimirov, N., L.~L\o{}vdok, D.~Lebiedz, and V.~Sourjik, 2008.
\newblock Dependence of bacterial chemotaxis on gradient shape and adaptation
  rate.
\newblock \emph{PLoS Comput. Biol.} 4:e1000242.

\bibitem[Bray et~al.(1998)Bray, Levin, and {Morton-Firth}]{BrayLevMFirth98}
Bray, D., M.~D. Levin, and C.~J. {Morton-Firth}, 1998.
\newblock Receptor clustering as a cellular mechanism to control sensitivity.
\newblock \emph{Nature} 393:85--88.

\bibitem[Duke and Bray(1999)]{DukeBray99}
Duke, T.~A.~J., and D.~Bray, 1999.
\newblock {Heightened sensitivity of a lattice of membrane receptors.}
\newblock \emph{Proc Natl Acad Sci U.S.A.} 96:10104--10108.

\bibitem[Mello and Tu(2005)]{MelTu05}
Mello, B.~A., and Y.~Tu, 2005.
\newblock An allosteric model for heterogeneous receptor complexes:
  understanding bacterial chemotaxis responses to multiple stimuli.
\newblock \emph{Proc Natl Acad Sci U.S.A.} 102:17354--17359.

\bibitem[Sourjik and Berg(2004)]{SouBerg04}
Sourjik, V., and H.~C. Berg, 2004.
\newblock Functional interactions between receptors in bacterial chemotaxis.
\newblock \emph{Nature} 428:437--441.

\bibitem[Keymer et~al.(2006)Keymer, Endres, Skoge, Meir, and
  Wingreen]{KeyEndSko06}
Keymer, J.~E., R.~G. Endres, M.~Skoge, Y.~Meir, and N.~S. Wingreen, 2006.
\newblock Chemosensing in Escherichia coli: two regimes of two-state receptors.
\newblock \emph{Proc. Natl. Acad. Sci. U.S.A.} 103:1786--1791.

\bibitem[Barkai and Leibler(1997)]{BarLei97}
Barkai, N., and S.~Leibler, 1997.
\newblock Robustness in simple biochemical networks.
\newblock \emph{Nature} 387:913--917.

\bibitem[Clausznitzer et~al.(2010)Clausznitzer, Oleksiuk, Lovdok, Sourjik, and
  Endres]{ClauOleEnd10}
Clausznitzer, D., O.~Oleksiuk, L.~Lovdok, V.~Sourjik, and R.~G. Endres, 2010.
\newblock Chemotactic response and adaptation dynamics in {Escherichia} coli.
\newblock \emph{PLoS Comput. Biol.} 6:e1000784.

\bibitem[Yuan and Berg(2013)]{YuanBerg13}
Yuan, J., and H.~C. Berg, 2013.
\newblock Ultrasensitivity of an Adaptive Bacterial Motor.
\newblock \emph{J. Mol. Biol.} 425:1760 -- 1764.

\bibitem[Mears et~al.(2014a) Mears, Koirala, Rao, Golding, and
  Chemla]{Chemla14Elife}
Mears, P.~J., S.~Koirala, C.~V. Rao, I.~Golding, and Y.~R. Chemla, 2014.
\newblock Escherichia coli swimming is robust against variations in flagellar
  number.
\newblock \emph{Elife} 3:e01916.

\bibitem[Lazova et~al.(2011)Lazova, Ahmed, Bellomo, Stocker, and
  Shimizu]{LazAhmShi2011}
Lazova, M.~D., T.~Ahmed, D.~Bellomo, R.~Stocker, and T.~S. Shimizu, 2011.
\newblock Response rescaling in bacterial chemotaxis.
\newblock \emph{Proc Natl Acad Sci U.S.A.} 108:13870--13875.

\bibitem[Sourjik and Berg(2002a)]{SouBerg02a}
Sourjik, V., and H.~C. Berg, 2002.
\newblock Receptor sensitivity in bacterial chemotaxis.
\newblock \emph{Proc. Natl. Acad. Sci. U.S.A.} 99:123--127.

\bibitem[Endres and Wingreen(2006)]{EndWin06}
Endres, R.~G., and N.~S. Wingreen, 2006.
\newblock Precise adaptation in bacterial chemotaxis through assistance
  neighborhoods.
\newblock \emph{Proc Natl Acad Sci U.S.A.} 103:13040--13044.

\bibitem[Sourjik and Berg(2002b)]{SourjikBerg2002b}
Sourjik, V., and H.~C. Berg, 2002.
\newblock Binding of the \emph{Escherichia coli} response regulator {CheY} to
  its target measured in vivo by fluorescence resonance energy transfer.
\newblock \emph{Proc Natl Acad Sci U.S.A} 99:12669--12674.

\bibitem[Neumann et~al.(2014)Neumann, Vladimirov, Krembel, Wingreen, and
  Sourjik]{Neumann14}
Neumann, S., N.~Vladimirov, A.~K. Krembel, N.~S. Wingreen, and V.~Sourjik,
  2014.
\newblock Imprecision of adaptation in Escherichia coli chemotaxis.
\newblock \emph{PloS One} 9:e84904.

\bibitem[Micali and Endres(2016)]{Micali16}
Micali, G., and R.~G. Endres, 2016.
\newblock Bacterial chemotaxis: information processing, thermodynamics, and
  behavior.
\newblock \emph{Curr. Opin. Microbiol.} 30:8 -- 15.

\bibitem[Lan and Tu(2016)]{TuLan16Review}
Lan, G., and Y.~Tu, 2016.
\newblock Information processing in bacteria: memory, computation, and
  statistical physics: a key issues review.
\newblock \emph{Reports on Progress in Physics} 79:052601.

\bibitem[De~Palo and Endres(2013)]{DePalo13}
De~Palo, G., and R.~G. Endres, 2013.
\newblock Unraveling adaptation in eukaryotic pathways: Lessons from
  protocells.
\newblock \emph{PLoS Comput. Biol.} 9:e1003300.

\bibitem[Barato et~al.(2014)Barato, Hartich, and Seifert]{Barato2014}
Barato, A.~C., D.~Hartich, and U.~Seifert, 2014.
\newblock Efficiency of cellular information processing.
\newblock \emph{New J. Phys.} 16:103024.

\bibitem[Govern and ten Wolde(2014)]{tenWolde14PNAS}
Govern, C.~C., and P.~R. ten Wolde, 2014.
\newblock Optimal resource allocation in cellular sensing systems.
\newblock \emph{Proc. Natl. Acad. Sci. U.S.A.} 111:17486--17491.

\bibitem[Monod et~al.(1965)Monod, Wyman, and Changeux]{MWC65}
Monod, J., J.~Wyman, and J.~P. Changeux, 1965.
\newblock ON THE NATURE OF ALLOSTERIC TRANSITIONS: A PLAUSIBLE MODEL.
\newblock \emph{J. Mol. Biol.} 12:88--118.

\bibitem[Cluzel et~al.(2000)Cluzel, Surette, and Leibler]{CluSurLei00}
Cluzel, P., M.~Surette, and S.~Leibler, 2000.
\newblock An ultrasensitive bacterial motor revealed by monitoring signaling
  proteins in single cells.
\newblock \emph{Science} 287:1652--1655.

\bibitem[Sneddon et~al.(2012)Sneddon, Pontius, and Emonet]{Sneddon12}
Sneddon, M.~W., W.~Pontius, and T.~Emonet, 2012.
\newblock Stochastic coordination of multiple actuators reduces latency and
  improves chemotactic response in bacteria.
\newblock \emph{Proc. Natl. Acad. Sci. U.S.A.} 109:805--810.

\bibitem[Mears et~al.(2014b)Mears, Koirala, Rao, Golding, and
  Chemla]{Chemla14}
Mears, P.~J., S.~Koirala, C.~V. Rao, I.~Golding, and Y.~R. Chemla, 2014.
\newblock Escherichia coli swimming is robust against variations in flagellar
  number.
\newblock \emph{Elife} 3:e01916.

\bibitem[Berg et~al.(1972)Berg, Brown, et~al.]{berg1972chemotaxis}
Berg, H.~C., D.~A. Brown, et~al., 1972.
\newblock Chemotaxis in Escherichia coli analysed by three-dimensional
  tracking.
\newblock \emph{Nature} 239:500--504.

\bibitem[Lovely and Dahlquist(1975)]{Lovely75}
Lovely, P.~S., and F.~Dahlquist, 1975.
\newblock Statistical measures of bacterial motility and chemotaxis.
\newblock \emph{J. Theor. Biol.} 50:477--496.

\bibitem[Endres et~al.(2008)Endres, Oleksiuk, Hansen, Meir, Sourjik, and
  Wingreen]{EndSouWin2008}
Endres, R.~G., O.~Oleksiuk, C.~H. Hansen, Y.~Meir, V.~Sourjik, and N.~S.
  Wingreen, 2008.
\newblock Variable sizes of \emph{Escherichia coli} chemoreceptor signaling
  teams.
\newblock \emph{Mol. Syst. Biol.} 4:211.

\bibitem[He et~al.(2016)He, Zhang, and Yuan]{Junhua16}
He, R., R.~Zhang, and J.~Yuan, 2016.
\newblock Noise-Induced Increase of Sensitivity in Bacterial Chemotaxis.
\newblock \emph{Biophysical Journal} 111:430--437.

\bibitem[Ahmed et~al.(2010)Ahmed, Shimizu, and Stocker]{Ahmed10}
Ahmed, T., T.~S. Shimizu, and R.~Stocker, 2010.
\newblock Microfluidics for bacterial chemotaxis.
\newblock \emph{Integr. Biol.} 2:604--629.

\bibitem[Crocker and Grier(1996)]{Crocker1996}
Crocker, J.~C., and D.~G. Grier, 1996.
\newblock Methods of digital video microscopy for colloidal studies.
\newblock \emph{J. Colloid Interface Sci.} 179:298--310.

\bibitem[Tostevin and ten Wolde(2009)]{TosttenWolde09}
Tostevin, F., and P.~R. ten Wolde, 2009.
\newblock Mutual information between input and output trajectories of
  biochemical networks.
\newblock \emph{Phys. Rev. Lett.} 102:218101--218101.

\bibitem[Brunel and Nadal(1998)]{BrunelNadal}
Brunel, N., and J.~P. Nadal, 1998.
\newblock Mutual information, Fisher information, and population coding.
\newblock \emph{Neural. Comput.} 10:1731--1757.

\bibitem[Qian and Beard(2005)]{Qian05}
Qian, H., and D.~A. Beard, 2005.
\newblock Thermodynamics of stoichiometric biochemical networks in living
  systems far from equilibrium.
\newblock \emph{Biophys. Chem.} 114:213--220.

\bibitem[Qian(2007)]{Qian07}
Qian, H., 2007.
\newblock Phosphorylation energy hypothesis: open chemical systems and their
  biological functions.
\newblock \emph{Annu. Rev. Phys. Chem.} 58:113--142.

\bibitem[Gabel and Berg(2003)]{Berg2003motors}
Gabel, C.~V., and H.~C. Berg, 2003.
\newblock The speed of the flagellar rotary motor of Escherichia coli varies
  linearly with protonmotive force.
\newblock \emph{Proc Natl Acad Sci U.S.A.} 100:8748--8751.

\bibitem[Li and Hazelbauer(2004)]{LiHaz04}
Li, M., and G.~L. Hazelbauer, 2004.
\newblock Cellular stoichiometry of the components of the chemotaxis signaling
  complex.
\newblock \emph{J. Bacteriol.} 186:3687--3694.

\bibitem[Sevick et~al.(2008)Sevick, Prabhakar, Williams, and Searles]{Sevick08}
Sevick, E., R.~Prabhakar, S.~R. Williams, and D.~J. Searles, 2008.
\newblock Fluctuation Theorems.
\newblock \emph{Annu. Rev. Phys. Chem.} 59:603--633.

\end{thebibliography}
\end{document}